\documentclass[iop, apj]{emulateapj}
\usepackage[]{hyperref}
\usepackage{textcomp}
\usepackage{placeins}
\usepackage{multirow}
\usepackage{graphicx}
\usepackage{amsmath}
\usepackage{amssymb,latexsym}
\usepackage{threeparttable}
\usepackage{longtable}
\usepackage{mhchem}
\def\jnl@style{\it}
\def\aaref@jnl#1{{\jnl@style#1}}

\def\aaref@jnl#1{{\jnl@style#1}}

\def\aj{\aaref@jnl{AJ}}                   
\def\araa{\aaref@jnl{ARA\&A}}             
\def\apj{\aaref@jnl{ApJ}}                 
\def\apjl{\aaref@jnl{ApJ}}                
\def\apjs{\aaref@jnl{ApJS}}               
\def\ao{\aaref@jnl{Appl.~Opt.}}           
\def\apss{\aaref@jnl{Ap\&SS}}             
\def\aap{\aaref@jnl{A\&A}}                
\def\aapr{\aaref@jnl{A\&A~Rev.}}          
\def\aaps{\aaref@jnl{A\&AS}}              
\def\azh{\aaref@jnl{AZh}}                 
\def\baas{\aaref@jnl{BAAS}}               
\def\jrasc{\aaref@jnl{JRASC}}             
\def\memras{\aaref@jnl{MmRAS}}            
\def\mnras{\aaref@jnl{MNRAS}}             
\def\pra{\aaref@jnl{Phys.~Rev.~A}}        
\def\prb{\aaref@jnl{Phys.~Rev.~B}}        
\def\prc{\aaref@jnl{Phys.~Rev.~C}}        
\def\prd{\aaref@jnl{Phys.~Rev.~D}}        
\def\pre{\aaref@jnl{Phys.~Rev.~E}}        
\def\prl{\aaref@jnl{Phys.~Rev.~Lett.}}    
\def\pasp{\aaref@jnl{PASP}}               
\def\pasj{\aaref@jnl{PASJ}}               
\def\jsara{\aaref@jnl{JSARA}}             
\def\qjras{\aaref@jnl{QJRAS}}             
\def\skytel{\aaref@jnl{S\&T}}             
\def\solphys{\aaref@jnl{Sol.~Phys.}}      
\def\sovast{\aaref@jnl{Soviet~Ast.}}      
\def\ssr{\aaref@jnl{Space~Sci.~Rev.}}     
\def\zap{\aaref@jnl{ZAp}}                 
\def\nat{\aaref@jnl{Nature}}              
\def\iaucirc{\aaref@jnl{IAU~Circ.}}       
\def\aplett{\aaref@jnl{Astrophys.~Lett.}} 
\def\apspr{\aaref@jnl{Astrophys.~Space~Phys.~Res.}}
\def\bain{\aaref@jnl{Bull.~Astron.~Inst.~Netherlands}} 
\def\fcp{\aaref@jnl{Fund.~Cosmic~Phys.}}  
\def\gca{\aaref@jnl{Geochim.~Cosmochim.~Acta}}   
\def\grl{\aaref@jnl{Geophys.~Res.~Lett.}} 
\def\jcp{\aaref@jnl{J.~Chem.~Phys.}}      
\def\jgr{\aaref@jnl{J.~Geophys.~Res.}}    
\def\jqsrt{\aaref@jnl{J.~Quant.~Spec.~Radiat.~Transf.}}
\def\memsai{\aaref@jnl{Mem.~Soc.~Astron.~Italiana}}
\def\nphysa{\aaref@jnl{Nucl.~Phys.~A}}   
\def\physrep{\aaref@jnl{Phys.~Rep.}}   
\def\physscr{\aaref@jnl{Phys.~Scr}}   
\def\planss{\aaref@jnl{Planet.~Space~Sci.}}   
\def\procspie{\aaref@jnl{Proc.~SPIE}}   
\def\nar{\aaref@jnl{NewAR}}   
\def\icarus{\aaref@jnl{Icarus}}

\usepackage{natbib}

\hypersetup{
     colorlinks   = true,
     citecolor     = blue
} 

\newcommand{\mos}{\,m\,s$^{-1}$}
\newcommand{\kms}{\,km\,s$^{-1}$}

\newcommand\msini{\ifmmode{{\mathrm M} \sin i}\else${{\mathrm M} \sin i}$\fi}

\slugcomment{Published in the Astrophysical Journal}

\shorttitle{Spin--orbit alignment of WASP-103\MakeLowercase{b}, WASP-87\MakeLowercase{b}, \& WASP-66\MakeLowercase{b}}
\shortauthors{Addison et al.}

\begin{document}

\title{Spin--orbit alignment for Three Transiting Hot Jupiters: WASP-103\MakeLowercase{b}, WASP-87\MakeLowercase{b}, \& WASP-66\MakeLowercase{b}\altaffilmark{$^{\dagger}$}}

\author{B. C. Addison \altaffilmark{1,2}, C. G. Tinney\altaffilmark{1,2}, D. J. Wright\altaffilmark{1,2}, D. Bayliss\altaffilmark{3}}

\email{baddison2005@gmail.com}

\altaffiltext{1}{Exoplanetary Science Group, School of Physics, University of New South Wales, Sydney, NSW 2052, Australia}
\altaffiltext{2}{Australian Centre of Astrobiology, University of New South Wales, Sydney, NSW 2052, Australia}
\altaffiltext{3}{Research School of Astronomy and Astrophysics, Australian National University, Canberra, ACT 2611, Australia}
\altaffiltext{$\dagger$}{Based on observations obtained at the Anglo-Australian Telescope, Siding Spring, Australia.}

\date{\today}

\begin{abstract}
We have measured the sky-projected spin--orbit alignments for three transiting Hot Jupiters, WASP-103b, WASP-87b, and WASP-66b, using spectroscopic measurements of the Rossiter--McLaughlin effect, with the CYCLOPS2 optical fiber bundle system feeding the UCLES spectrograph on the Anglo-Australian Telescope. The resulting sky-projected spin--orbit angles of $\lambda = 3^{\circ}\pm33^{\circ}$, $\lambda = -8^{\circ}\pm11^{\circ}$, and $\lambda = -4^{\circ}\pm22^{\circ}$ for WASP-103b, WASP-87b, and WASP-66b, respectively, suggest that these three planets are likely on nearly aligned orbits with respect to their host star's spin axis. WASP-103 is a particularly interesting system as its orbital distance is only 20\% larger than its host star's Roche radius and the planet likely experiences strong tidal effects. WASP-87 and WASP-66 are hot ($T_\mathrm{{eff}}=6450\pm120$\,K and $T_\mathrm{{eff}}=6600\pm150$\,K, respectively) mid-F stars, making them similar to the majority of stars hosting planets on high-obliquity orbits. Moderate spin--orbit misalignments for WASP-103b and WASP-66b are consistent with our data, but polar and retrograde orbits are not favored for these systems.
\end{abstract}


\section{INTRODUCTION}
\setcounter{footnote}{3}
Measurements of the projected obliquity (i.e., sky-projected angle between planetary orbits and their host star's spin axis) of exoplanetary systems are key to understanding the various mechanisms involved in the formation and migration of extrasolar planets \citep[e.g.,][]{2012ApJ...757...18A}. As of 2015 November, 91 exoplanetary systems\footnote{This study has made use of Ren\'{e} Heller's Holt-Rossiter--McLaughlin Encyclopaedia and was last updated on 2015 November \url{http://www.astro.physik.uni-goettingen.de/~rheller/}.}, including WASP-66, WASP-87, and WASP-103 as reported here, have measured projected obliquities. These measurements have revealed a stunning diversity of planetary orbits that includes 36 planets on significantly misaligned orbits ($|\lambda| > 22.5^{\circ}$), 15 of which are on nearly polar orbits ($67.5^{\circ} < |\lambda| < 112.5^{\circ}$ or $247.5^{\circ} < |\lambda| < 292.5^{\circ}$), and 9 are on retrograde orbits ($112.5^{\circ} \leq |\lambda| \leq 247.5^{\circ}$). The vast majority of reported spin--orbit alignments come from spectroscopic measurements of the Rossiter--McLaughlin effect \citep[e.g.,][]{1924ApJ....60...22M,1924ApJ....60...15R,2000A&A...359L..13Q,2005ApJ...622.1118O}, a radial velocity anomaly produced during planetary transits from the rotationally broadened stellar line profiles of a star being asymmetrically distorted when specific regions of the stellar disk are occulted by a transiting planet.

Hot Jupiters orbiting stars cooler than $6250$\,K have been observed to be generally in spin--orbit alignment, while hotter stars are seen to host high-obliquity systems, as noted by \citet{2010ApJ...718L.145W}, \citet{2012ApJ...757...18A}, and others. The reason for this observed dichotomy is thought to be linked to the amount of mass in the stellar convective envelope, which acts to tidally dampen orbital obliquities. Therefore, the realignment timescale for planets is believed to be correlated with the stellar convective envelope mass. Cooler stars have a thicker convective envelope than hotter stars \citep[as supported by stellar interior models; see][]{2001ApJ...556L..59P} and can thus drive planets that are on highly misaligned orbits onto low-obliquity orbits more quickly. \citet{2012ApJ...757...18A} proposes that the mechanism(s) responsible for migrating giant planets into short period orbits are also randomly misaligning their orbits. Stars with $T_\mathrm{{eff}} > 6250$\,K can only weakly dampen orbital obliquities and are thus unable to realign planetary orbits. Therefore, the \citet{2012ApJ...757...18A} model predicts stars with $T_\mathrm{{eff}} > 6250$\,K should be observed to host planets on a random distribution of orbital obliquities while cooler stars should host planets on nearly aligned orbits.

An expansion of the parameter space for which spin--orbit angles are measured will be important for testing whether these observed trends continue to hold and so test models for orbital migration. In particular, obliquity measurements need to be carried out for suitable systems that belong to the least explored parameter space, which includes sub-Jovian, long-period, and multi-planet systems. Several mechanisms have been proposed for producing Hot Jupiters and misaligning their orbits and these can generally be grouped into two categories: disk migration and high eccentricity migration. Disk migration occurs through the interactions between a planet and its surrounding protoplanetary disk \citep[e.g., see][]{1997Icar..126..261W}. Migration through this process leads to the production of short-period planets on well aligned orbits \citep{2010MNRAS.401.1505B}; therefore, this mechanism is disfavored for producing the observed population of Hot Jupiters on high-obliquity orbits. High eccentricity migration through either Kozai-Lidov resonances \citep{1962AJ.....67..591K,1962P&SS....9..719L,2007ApJ...669.1298F}, planet-planet scatterings \citep{2008ApJ...686..580C}, secular chaos \citep{2011ApJ...735..109W}, or some combination of these mechanisms appears to be the likely route for producing misaligned Hot Jupiters.

To further expand the sample of planets with spin--orbit alignment measurements, we have observed the Rossiter--McLaughlin effect for WASP-66b, WASP-87b, and WASP-103b, three recently discovered Hot Jupiter planets from the Wide Angle Search for Planets \citep[see][]{2012MNRAS.426..739H,2014arXiv1410.3449A,2014A&A...562L...3G}. These systems were predicted to have large observable velocity anomalies and were good candidates for follow up observations to determine their orbital obliquities. 

WASP-103 is a late F star with a mass of $M_{\star}=1.220^{+0.039}_{-0.036}$\,$M_{\odot}$, a radius of $R_{\star}=1.436^{+0.052}_{-0.031}$\,$R_{\odot}$, an effective temperature of $T_\mathrm{{eff}}=6110\pm160$\,K, and has a moderate rotation ($v \sin i_{\star}=10.6\pm0.9$\,\kms) as reported by \citet{2014A&A...562L...3G}. It hosts a planet with a mass of $M_{P}=1.490\pm0.088$\,$M_{J}$, moderately inflated with a radius of $R_{P}=1.528^{+0.073}_{-0.041}$\,$R_{J}$, and an orbital period of just $P=0.925542\pm0.000019$\,day \citep{2014A&A...562L...3G}. 

WASP-103 is a particularly interesting system as it consists of a Hot Jupiter that is orbiting at only 1.2 times the Roche radius of the host star and 1.5 times its stellar diameter \citep{2014A&A...562L...3G}. The planet likely experiences strong tidal forces that cause significant mass loss with Roche-lobe overflow and is very near the edge of being tidally disrupted. Measuring the spin--orbit angle for this system could potentially offer insights into the processes involved in the migration of WASP-103b to its current ultra-short-period orbit. This planet currently has the second shortest orbital period of all planetary systems with reported spin--orbit angle measurements \citep[WASP-19b has the shortest orbital period with a measured obliquity;][]{2011ApJ...730L..31H,2012ApJ...757...18A,2013MNRAS.428.3671T}.

WASP-87 is a mid-F star with a mass of $M_{\star}=1.204\pm0.093$\,$M_{\odot}$, a radius of $R_{\star}=1.627\pm0.062$\,$R_{\odot}$, an effective temperature of $T_\mathrm{{eff}}=6450\pm120$\,K, and rotating with a $v \sin i_{\star}=9.6\pm0.7$\,\kms, as reported in \citet{2014arXiv1410.3449A}. It hosts a giant planet with a mass of $M_{P}=2.18\pm0.15$\,$M_{J}$, with a radius of $R_{P}=1.385\pm0.060$\,$R_{J}$, and an orbital period of $P=1.6827950\pm0.0000019$\,day \citep{2014arXiv1410.3449A}. A possible bound early-G stellar companion was observed $8.2^{\texttt{"}}$ from WASP-87 by \citet{2014arXiv1410.3449A}. WASP-87 was predicted to be a good candidate for follow up Rossiter--McLaughlin observations due to the high $v \sin i_{\star}$ and large $R_{P}$. 

WASP-66 is a mid-F star with a mass of $M_{\star}=1.30\pm0.07$\,$M_{\odot}$, a radius of $R_{\star}=1.75\pm0.09$\,$R_{\odot}$, effective temperature of $T_\mathrm{{eff}}=6600\pm150$\,K, and rotating with a $v \sin i_{\star}=13.4\pm0.9$\,\kms, as reported in \citet{2012MNRAS.426..739H}. It hosts a massive planet with a mass of $M_{P}=2.32\pm0.13$\,$M_{J}$, slightly inflated with a radius of $R_{P}=1.39\pm0.09$\,$R_{J}$, and an orbital period of $P=4.086052\pm0.000007$\,day \citep{2012MNRAS.426..739H}. WASP-66 was also predicted to be a good candidate for follow up Rossiter--McLaughlin observations. 

\section{OBSERVATIONS}
We carried out the spectroscopic observations of WASP-103b, WASP-87b, and WASP-66b using the CYCLOPS2 fiber feed with the UCLES spectrograph on the Anglo-Australian Telescope (AAT). The instrumental set up and observing strategy for the transit observations closely followed that presented in our previous Rossiter--McLaughlin publications \citep[i.e., WASP-79b and HATS-3b;][]{2013ApJ...774L...9A,2014ApJ...792..112A}. We used a thorium--argon calibration lamp (ThAr) to illuminate all on-sky fibers, and a thorium--uranium--xenon lamp (ThUXe) to illuminate the simultaneous calibration fiber for calibrating the observations. We provide a summary of the observations in Tables\,\ref{table:spectro_obs_wasp-103} \& \ref{table:spectro_obs_wasp-66_87}.

\begin{table}
\centering
\begin{minipage}{\columnwidth}
\caption{Summary of WASP-103b Transit Spectroscopic Observations.}
\centering
\resizebox{\columnwidth}{!}{%
\begin{tabular}{l c c}
\hline\hline \\ [-2.0ex]
 & WASP-103b (obs 1) & WASP-103b (obs 2) \\ [0.5ex]
\hline \\ [-2.0ex]
UT Time of Obs & 14:48-15:08\,UT & 14:27-18:22\,UT \\
UT Date of Obs & 2014 May 21 & 2014 May 22 \\
Cadence & 1175\,s & 1175\,s \\
Readout Times & 175\,s & 175\,s \\
Readout Speed & Normal & Normal \\
Readout Noise & 3.19\,$e^{-}$ & 3.19\,$e^{-}$ \\
S/N (/2.5~pix at $\lambda=5490$~\AA{}) & 27-29 & 27-29 \\
Resolution (${\lambda}/{\Delta}{\lambda}$) & 70,000 & 70,000 \\
Number of Spectra & 2 & 13 \\
Seeing & $1.0^{\texttt{"}}$ & $0.8^{\texttt{"}}$-$1.4^{\texttt{"}}$ \\
Weather Conditions & Clear & Some clouds \\
Airmass Range & 1.3 & 1.3-2.4 \\
\hline \\ [-2.5ex]
\end{tabular}}
\label{table:spectro_obs_wasp-103}
\end{minipage}
\end{table}

\begin{table}
\centering
\begin{minipage}{\columnwidth}
\caption{Summary of WASP-87b and WASP-66b Transit Spectroscopic Observations.}
\centering
\resizebox{\columnwidth}{!}{%
\begin{tabular}{l c c}
\hline\hline \\ [-2.0ex]
 & WASP-87b & WASP-66b \\ [0.5ex]
\hline \\ [-2.0ex]
UT Time of Obs & 11:03-16:44\,UT & 12:25-17:46\,UT \\
UT Date of Obs & 2015 Feb 28 & 2014 Mar 21 \\
Cadence & 1275\,s & 1375\,s \\
Readout Times & 175\,s & 175\,s \\
Readout Speed & Normal & Normal \\
Readout Noise & 3.19\,$e^{-}$ & 3.19\,$e^{-}$ \\
S/N (/2.5~pix at $\lambda=5490$~\AA{}) & N/A & 37--39 \\
Resolution (${\lambda}/{\Delta}{\lambda}$) & 70,000 & 70,000 \\
Number of Spectra & 17 & 15 \\
Seeing & $1.0^{\texttt{"}}$ & $1.2^{\texttt{"}}$ \\
Weather Conditions & Some clouds & Some clouds \\
Airmass Range & 1.1-1.8 & 1.0-2.2 \\
\hline \\ [-2.5ex]
\end{tabular}}
\label{table:spectro_obs_wasp-66_87}
\end{minipage}
\end{table}

\subsection{Spectroscopic Observations of WASP-103b}
Spectroscopic transit observations of WASP-103b were obtained on the night of 2014 May 22, starting $\sim\!50$\,minutes before ingress and finishing $\sim\!15$\,minutes after egress. A total of 13 spectra were obtained on that night (7 during the $\sim\!2.5$\,hr transit) in good observing conditions for Siding Spring Observatory with seeing between $0.7^{\texttt{"}}$ and $1.1^{\texttt{"}}$ and some patchy clouds. WASP-103 was observed at an airmass of 1.3 for the first exposure, 1.5 near mid-transit, and 2.4 at the end of the observations. A $\mathrm{S/N}=29$ per 2.5 pixel resolution element at $\lambda=5490$\,\AA{} (in total over all 16 fibers) was obtained at an airmass of 1.3, $0.8^{\texttt{"}}$ seeing, and integration times of 1000\,s.

We also obtained two out-of-transit observations of WASP-103b on the previous night (May 21) and attempted to use them to determine the radial velocity offset between our data set and the \citet{2014A&A...562L...3G} data set. Observing conditions on this night were good with seeing $\sim\!1.0^{\texttt{"}}$ and clear skies. A $\mathrm{S/N}=28$ per 2.5 pixel resolution element at $\lambda=5490$\,\AA{} (in total over all 16 fibers) was obtained for WASP-103 when observed at an airmass of 1.3.

\subsection{Spectroscopic Observations of WASP-87b}
We observed WASP-87b on the night of 2015 February 28, starting 70\,minutes before ingress and finishing 80\,minutes after egress. A total of 17 spectra were obtained on that night (including 9 during the $\sim\!3$\,hr transit) in good observing conditions with seeing between $0.9^{\texttt{"}}$ and $1.3^{\texttt{"}}$. WASP-87 was observed at an airmass of 1.8 for the first exposure, 1.20 near mid-transit, and 1.1 at the end of the observations. 

\subsection{Spectroscopic Observations of WASP-66b}
We obtained transit observations of WASP-66b on the night of 2014 March 21, starting $\sim\!30$\,minutes before ingress and finishing $\sim\!20$\,minutes after egress (see Table\,\ref{table:spectro_obs_wasp-66_87} for a summary of these observations). A total of 15 spectra with an exposure time of 1200\,s were obtained on that night (11 during the $\sim\!4.5$\,hr transit) in good observing conditions with seeing $\sim\!1.2^{\texttt{"}}$ and some patchy clouds. The airmass at which WASP-66 was observed varied from 1.0 for the first exposure, 1.1 near mid-transit, and 2.2 at the end of the observations. A $\mathrm{S/N}=39$ per 2.5 pixel resolution element at $\lambda=5490$\,\AA{} (in total over all 16 fibers) was obtained at an airmass of 1.0 and in $1.2^{\texttt{"}}$ seeing.

\subsection{Independent Determination of Stellar Rotational Velocity} \label{sec:vsini}
We determined the stellar rotational velocity for WASP-103 and WASP-66 independently of the Rossiter--McLaughlin effect by fitting a rotationally broadened Gaussian to a least-squares deconvolution line profile for every spectral order \citep[as done in][]{2013ApJ...774L...9A,2014ApJ...792..112A} of the two best spectra of WASP-103 and WASP-66. For WASP-103, $v\sin i_{\star}=8.8 \pm 0.7$\,\kms, and for WASP-66, $v\sin i_{\star}=11.8 \pm 0.4$\,\kms. $v\sin i_{\star}$ determined from the Rossiter--McLaughlin effect for these two systems (as presented in \autoref{sec:Discussion_wasp-66-103}) is consistent with the values determined from the least-squares deconvolution method, but with significantly larger uncertainties. $v\sin i_{\star}$ as reported in \citet{2014A&A...562L...3G} and \citet{2012MNRAS.426..739H} for WASP-103 and WASP-66, respectively, are inconsistent with the values determined from the least-squares deconvolution method. \citet{2014A&A...562L...3G} did not specify the method they used to derive the $v\sin i_{\star}$ value of WASP-103. \citet{2012MNRAS.426..739H} and \citet{2014arXiv1410.3449A} determined $v\sin i_{\star}$ values of WASP-66 and WASP-87, respectively, by fitting several unblended Fe \ce{I} line profiles.

\section{Rossiter--McLaughlin Analysis} \label{sec:RM_analysis}

Spectroscopic data were reduced by tracing each fiber and optimally extracting each spectral order using custom MATLAB routines developed by the authors \citep[see][]{2013ApJ...774L...9A,2014ApJ...792..112A}. For WASP-103 and WASP-66, each of the 15 useful fibers, in each of the 17 useful orders, are used to estimate a radial velocity (and associated uncertainty) by cross-correlation with a spectrum of a bright template star, HD 157347, of similar spectral type to the targets, using the IRAF task, \textit{fxcor}, as described in \citet{2013ApJ...774L...9A,2014ApJ...792..112A}. For WASP-87, we found the best radial velocities were produced by cross-correlation of each of the 15 useful fibers in each of the 17 useful orders with a 5000\,K synthetic template star. A variety of templates were trialed for cross-correlation, including observations of other bright template stars (such as HD 10700, HD 206395, and HD 86264), as well as synthetic spectra of F- and G-type stars. The lowest inter-fider\footnote{The term `fider' refers to the spectrum extracted from a single fiber in a single spectral order in the echellogram.} velocity scatter was obtained using the spectrum of HD 157347 for WASP-103 and WASP-66. The weighted average velocities for each observation were computed and the uncertainties for each weighted velocity were estimated from the weighted standard deviation of the fider velocity scatter. The weighted radial velocities for the WASP-103, WASP-87, and WASP-66 transit observations, including their uncertainties and total signal-to-noise ratio (S/N), are given in Tables \ref{table:WASP-103_RVs}, \ref{table:WASP-87_RVs}, and \ref{table:WASP-66_RVs}, respectively.

The Exoplanetary Orbital Simulation and Analysis Model \citep[ExOSAM; see][]{2013ApJ...774L...9A,2014ApJ...792..112A} was used to determine the best fit $\lambda$ and $v\sin i_{\star}$ values for WASP-103, WASP-87, and WASP-66 from Rossiter--McLaughlin effect measurements. We have implemented a Metropolis--Hastings Markov Chain Monte Carlo (MCMC) algorithm in ExOSAM that replaces the Monte Carlo model used in \citet{2014ApJ...792..112A} to derive accurate posterior probability distributions of $\lambda$ and $v\sin i_{\star}$ and to opitmize their fit to the radial velocity data. Our MCMC procedure largely follows from \citet{2007MNRAS.380.1230C} and is outlined as follows. There are 16 input parameters used to model the Rossiter--McLaughlin effect for these three systems, of which 14 are prior values given by \citet{2014A&A...562L...3G} for WASP-103, \citet{2014arXiv1410.3449A} for WASP-87, and \citet{2012MNRAS.426..739H} for WASP-66. The 14 priors are: the planet-to-star radius ratio ($R_{p}/R_{\star}$); the orbital inclination angle ($I$); the orbital period ($P$); the mid-transit time ($T_{0}$) at the epoch of observation; a radial velocity offset ($V_{d}$) between the AAT data sets presented here and previously published data sets; a velocity offset term ($V_{s}$) accounting for systematic effects between the AAT data sets taken over multiple nights; planet-to-star mass ratio ($M_{p}/M_{\star}$); orbital eccentricity ($e$); argument of periastron ($\varpi$); two adopted quadratic limb-darkening coefficients ($q_{1}$ and $q_{2}$); the micro-turbulence velocity ($\xi_{t}$); the macro-turbulence velocity ($v_\mathrm{mac}$); and the center-of-mass velocity ($V_{T_{P}}$) at published epoch $T_{P}$. 

We fixed $q_{1}$ and $q_{2}$, $\xi_{t}$, $v_\mathrm{mac}$, and $V_{T_{P}}$ to their literature values. All three planets are also consistent with being on circular orbits so we fixed $e=0$ and $\varpi=0$. The two out-of-transit observations obtained on 2014 May 21 for WASP-103 could not be reliably used in constraining $V_{d}$ between our data and the \citet{2014A&A...562L...3G} data. This was due to systematic velocity offsets between the two nights that could not be well characterized from the small number of out-of-transit radial velocities that were obtained. We discarded the two observations from May 21 and determined $V_{d}$ using the six out-of-transit radial velocities taken on the night of the transit. Therefore, the velocity difference, $V_{s}$, between the May 21 and 22 data sets was not used in modeling the velocity anomaly. For WASP-66 and WASP-87, $V_{s}$ was not used in modeling the velocity anomaly since the objects were observed only on one night.

We assumed Gaussian distributions for the other six priors ($R_{p}/R_{\star}$, $I$, $P$, $T_{0}$, $V_{d}$, and $M_{p}/M_{\star}$) and allowed them to perform a random walk in the MCMC. Values for these parameters are randomly drawn from a Gaussian distribution centered on the previously accepted MCMC iteration and their reported $1\sigma$ uncertainties, as given in Tables \ref{table:WASP-103_Parameters}, \ref{table:WASP-87_Parameters}, and \ref{table:WASP-66_Parameters} for WASP-103, WASP-87, and WASP-66, respectively, as described by Equations \ref{equat:Rp}, \ref{equat:Inc}, \ref{equat:Period}, \ref{equat:transit_time}, \ref{equat:velocity_offset}, \& \ref{equat:Mp}. The $1\sigma$ prior uncertainties remained fixed in the MCMC, but are multiplied by the scale factor $f$, an adaptive step-size controller that evolves with the estimated uncertainties for the proposal parameters $\lambda$ and $v\sin i_{\star}$ as described below.

\begin{equation}\label{equat:Rp}
{R_{p}/R_{\star}}_{i} = {R_{p}/R_{\star}}_{i-1} + \sigma_{R_{p}/R_{\star}}G(0,1)f
\end{equation}
\begin{equation}\label{equat:Inc}
{I}_{i} = {I}_{i-1} + \sigma_{I}G(0,1)f
\end{equation}
\begin{equation}\label{equat:Period}
{P}_{i} = {P}_{i-1} + \sigma_{P}G(0,1)f
\end{equation}
\begin{equation}\label{equat:transit_time}
{T_{0}}_{i} = {T_{0}}_{i-1} + \sigma_{T_{0}}G(0,1)f
\end{equation}
\begin{equation}\label{equat:velocity_offset}
{V_{d}}_{i} = {V_{d}}_{i-1} + \sigma_{V_{d}}G(0,1)f
\end{equation}
\begin{equation}\label{equat:Mp}
{M_{p}/M_{\star}}_{i} = {M_{p}/M_{\star}}_{i-1} + \sigma_{M_{p}/M_{\star}}G(0,1)f
\end{equation}

where $i$ is the current iteration and $i-1$ is previously accepted MCMC iteration, $\sigma_{R_{p}/R_{\star}}$ is the standard deviation on ${R_{p}/R_{\star}}$, $\sigma_{I}$ is the standard deviation on $I$, $\sigma_{P}$ is the standard deviation on $P$, $\sigma_{T_{0}}$ is the standard deviation on $T_{0}$, $\sigma_{V_{d}}$ is the standard deviation on $V_{d}$, $\sigma_{M_{p}/M_{\star}}$ is the standard deviation on ${M_{p}/M_{\star}}$, $G(0,1)$ is a random Gaussian deviate of mean zero and standard deviation of unity, and $f$ is an adaptable scale factor that is used to ensure that the acceptance rate is maintained close to the optimal value of $25\%$ \citep[see][]{2004PhRvD..69j3501T,2007MNRAS.380.1230C}.

$\lambda$ and $v\sin i_{\star}$ represent our proposal parameters for describing the Rossiter--McLaughlin effect. They perform a random walk through the parameter space that maps out the joint posterior probability distribution by the generation of a cloud of points. A value is drawn for $\lambda$ and $v\sin i_{\star}$ at each MCMC iteration ($i$) by perturbing the previously accepted proposal values by a small random amount as described by Equations \ref{equat:lambda} and \ref{equat:vsini}.

\begin{equation}\label{equat:lambda}
\lambda_{i} = \lambda_{i-1} + \sigma_{\lambda}G(0,1)f
\end{equation}

\begin{equation}\label{equat:vsini}
v\sin i_{\star,i} = v\sin i_{\star,i-1} + \sigma_{v\sin i_{\star}}G(0,1)f
\end{equation}

where $\sigma_{\lambda}$ is the standard deviation on $\lambda$ and $\sigma_{v\sin i_{\star}}$ is the standard deviation on $v\sin i_{\star}$. 

We initially set $f=0.5$, but allow it to evolve in conjunction with the estimated proposal uncertainties. Following the procedure of \citet{2007MNRAS.380.1230C}, we adjust $f$ and update the uncertainties for $\lambda$ and $v\sin i_{\star}$ from the Markov chains themselves on every 100th accepted MCMC iteration. The reason for allowing the $\lambda$ and $v\sin i_{\star}$ uncertainties to evolve with $f$ is to ensure that their step sizes are sufficient for the adequate exploration of the parameter space. The simple linear algorithm $f_{new}=400f_{old}/N_{T}$ is used to determine $f$, where $N_{T}$ is the number of trailed proposals during the previous 100 successful iterations.

The best fit values and confidence intervals for $\lambda$ and $v\sin i_{\star}$ are determined from calculating the joint posterior probability distribution, given as the following:

\begin{multline}\label{equat:likelihood}
\mathcal{P}(R_{p}/R_{\star},I,P,T_{0},V_{d},M_{p}/M_{\star},\lambda,v\sin i_{\star}) \\ 
\mathcal{P}(D \mid \lambda, v\sin i_{\star}, R_{p}/R_{\star},I,P,T_{0},V_{d},M_{p}/M_{\star}) \\
=\exp(-\chi^{2}/2)
\end{multline}

Where $\mathcal{P}(R_{p}/R_{\star},I,P,T_{0},V_{d},M_{p}/M_{\star},\lambda,v\sin i_{\star})$ and $\mathcal{P}(D \mid \lambda, v\sin i_{\star}, R_{p}/R_{\star},I,P,T_{0},V_{d},M_{p}/M_{\star})$ are the prior probability distribution and the likelihood of obtaining the observed data $D$ given the model $M$, respectively. $\chi^{2}$ is calculated as:
\begin{multline}\label{equat:chi_squared}
\chi_{i}^{2}=\sum_{n=1}^{N_{D}}\left( \frac{\left\lgroup M_{n}-D_{n}\right\rgroup}{\sigma_{D_{n}}}\right)_{i}^{2}\\ 
+\frac{\left\lgroup (R_{p}/R_{\star})_{i}-(R_{p}/R_{\star})_{0}\right\rgroup^{2}}{\sigma_{R_{p}/R_{\star}}^{2}}
+\frac{\left\lgroup I_{i}-I_{0}\right\rgroup^{2}}{\sigma_{I}^{2}} \\
+\frac{\left\lgroup P_{i}-P_{0}\right\rgroup^{2}}{\sigma_{P}^{2}}
+\frac{\left\lgroup (T_{0})_{i}-(T_{0})_{0}\right\rgroup^{2}}{\sigma_{T_{0}}^{2}} 
+\frac{\left\lgroup (V_{d})_{i}-(V_{d})_{0}\right\rgroup^{2}}{\sigma_{V_{d}}^{2}} \\ 
+\frac{\left\lgroup (M_{p}/M_{\star})_{i}-(M_{p}/M_{\star})_{0}\right\rgroup^{2}}{\sigma_{M_{p}/M_{\star}}^{2}} \\
+\frac{\left\lgroup (v\sin i_{\star})_{i}-(v\sin i_{\star})_{0}\right\rgroup^{2}}{2\sigma_{v\sin i_{\star}}^{2}}
\end{multline}

where $i$ is the ith accepted proposal, $N_{D}$ is the number of radial velocities $D$ being fitted to the model $M$, and $n$ is the nth radial velocity datum.

To determine the best fit $\lambda$ for WASP-103, WASP-87, and WASP-66, we imposed a weak prior on $v\sin i_{\star}$ and assume a flat prior on $\lambda$. This was done by setting the prior to the $v\sin i_{\star}$ value determined from the least-squares deconvolution method for WASP-103 and WASP-66 (see Section \ref{sec:vsini}) and the value determined spectroscopically by \citet{2014arXiv1410.3449A} for WASP-87, using the $2\sigma_{v\sin i_{\star}}$ as the uncertainty for $v\sin i_{\star}$.  The decision to impose a weak prior on $v\sin i_{\star}$ for these three systems was based on reducing the potential bias on $\lambda$ from the large radial velocity uncertainties, inadequate sampling of the Rossiter--McLaughlin effect during their transits, and the strong correlations between $\lambda$ and $v\sin i_{\star}$ from the small impact parameter. 


The decision as to whether to accept or reject a given set of proposals is made by the Metropolis--Hastings rule \citep{2007MNRAS.380.1230C}. This rule states that if $Q_{i}\le Q_{i-1}$, then the new proposal values are accepted; otherwise if $Q_{i}>Q_{i-1}$, then the proposals are accepted with probability $\exp(-\Delta Q/2)$, where $\Delta Q\equiv Q_{i}-Q_{i-1}$ and $Q_{i}=\chi_{i}^{2}$. The algorithm first converges to the optimal solution and then explores the parameter space around it. 

The optimal solutions of $\lambda$ and $v\sin i_{\star}$ for WASP-103, WASP-87, and WASP-66 were computed from the mean of the MCMC chains. Likewise, $\sigma_{\lambda}$ and $\sigma_{v\sin i_{\star}}$ are computed from the standard deviation of their mean. We obtained sufficient convergence and well mixing of the Markov chains from 10,000 accepted MCMC iteration with no `burn-in' period. Burn-in is a colloquial term that describes the process in which a certain number of iterations at the start of an MCMC run are discarded and the rest are kept for calculating the best fit parameters and confidence intervals. \cite{brooks2011handbook} has suggested this procedure is mostly unnecessary as long as the Markov chain is started reasonably close to the equilibrium distribution (determined from preliminary MCMC runs or some prior knowledge of the distribution). Therefore, we follow the advice of \cite{brooks2011handbook} by not applying a burn-in phase in ExOSAM.

\begin{table}
\centering
\caption{Radial velocities for WASP-103 (fiber and order averaged) taken on 2014 May 21 and 22}
\label{table:WASP-103_RVs}
\centering
\resizebox{\columnwidth}{!}{%
\begin{threeparttable}[b]
\begin{tabular}{l c c c c l c c c}
\hline\hline \\ [-2.0ex]
Time & RV & S/N at & In/Out &  & Time & RV & S/N at & In/Out \\
BJD-2400000 & (\mos) & $\lambda$=5490\AA{} & Transit &  & BJD-2400000 & (\mos) & $\lambda$=5490\AA{} & Transit \\ [0.5ex]
\hline \\ [-2.0ex]
56458.62090\tnote{a} & -42466 $\pm$ 41 & N/A & Out &     \hspace{10 mm}		& 56800.15815 & -42520 $\pm$ 59 & 29 & In \\
56444.76341\tnote{a} & -42493 $\pm$ 45 & N/A & In &     \hspace{10 mm}		& 56800.17333 & -42519 $\pm$ 45 & 30 & In \\
56457.73633\tnote{a} & -42586 $\pm$ 47 & N/A & In &     \hspace{10 mm}		& 56800.18921 & -42564 $\pm$ 52 & 29 & In \\
56510.54044\tnote{a} & -42743 $\pm$ 38 & N/A & In &     \hspace{10 mm}		& 56800.20168 & -42702 $\pm$ 68 & 29 & In \\
56799.12290 & -42479 $\pm$ 52 & 29 & Out &     \hspace{10 mm}		& 56800.21947 & -42767 $\pm$ 58 & 28 & In \\
56799.13655 & -42478 $\pm$ 49 & 28 & Out &     \hspace{10 mm}		& 56800.23195 & -42871 $\pm$ 60 & 29 & In \\
56800.10822 & -42491 $\pm$ 71 & 30 & Out &     \hspace{10 mm}		& 56800.24443 & -42744 $\pm$ 59 & 29 & In \\
56800.12070 & -42542 $\pm$ 60 & 29 & Out &     \hspace{10 mm}		& 56800.25691 & -42766 $\pm$ 74 & 27 & Out \\
56800.13319 & -42529 $\pm$ 54 & 30 & Out &     \hspace{10 mm}		& 56800.27112 & -42780 $\pm$ 64 & 29 & Out \\
56800.14567 & -42547 $\pm$ 52 & 29 & Out &     \hspace{10 mm}		&  &  &  &  \\
\hline \\ [-2.5ex]
\end{tabular} 
\begin{tablenotes}
\item [a] \textit{Published near-transit radial velocities from \citet{2014A&A...562L...3G} that have been adjusted for the velocity offset between the data sets.}
\end{tablenotes}
\end{threeparttable}}
\end{table}

\begin{table}
\centering
\caption{Radial velocities for WASP-87 (fiber and order averaged) taken on 2015 February 28}
\label{table:WASP-87_RVs}
\centering
\resizebox{\columnwidth}{!}{%
\begin{threeparttable}[b]
\begin{tabular}{l c c c c l c c c}
\hline\hline \\ [-2.0ex]
Time & RV & S/N at & In/Out &  & Time & RV & S/N at & In/Out \\
BJD-2400000 & (\mos) & $\lambda$=5490\AA{} & Transit &  & BJD-2400000 & (\mos) & $\lambda$=5490\AA{} & Transit \\ [0.5ex]
\hline \\ [-2.0ex]
56361.75702\tnote{a} & -13323 $\pm$ 21 & N/A & Out &     \hspace{10 mm}		& 57082.07871 & -13416 $\pm$ 23 & N/A & In \\
56361.77401\tnote{a} & -13332 $\pm$ 22 & N/A & In &     \hspace{10 mm}		& 57082.09350 & -13475 $\pm$ 15 & N/A & In \\
57081.96030 & -13315 $\pm$ 23 & N/A & Out &     \hspace{10 mm}		& 57082.10831 & -13558 $\pm$ 16 & N/A & In \\
57081.97511 & -13269 $\pm$ 37 & N/A & Out &     \hspace{10 mm}		& 57082.12311 & -13561 $\pm$ 24 & N/A & In \\
57081.98990 & -13306 $\pm$ 29 & N/A & Out &     \hspace{10 mm}		& 57082.13790 & -13527 $\pm$ 25 & N/A & In \\
57082.00470 & -13344 $\pm$ 24 & N/A & Out &     \hspace{10 mm}		& 57082.15271 & -13540 $\pm$ 18 & N/A & Out \\
57082.01949 & -13328 $\pm$ 27 & N/A & In &     \hspace{10 mm}		& 57082.16751 & -13529 $\pm$ 29 & N/A & Out \\
57082.03430 & -13343 $\pm$ 22 & N/A & In &     \hspace{10 mm}		& 57082.18230 & -13507 $\pm$ 18 & N/A & Out \\
57082.04910 & -13312 $\pm$ 34 & N/A & In &     \hspace{10 mm}		& 57082.19711 & -13526 $\pm$ 23 & N/A & Out \\
57082.06391 & -13362 $\pm$ 28 & N/A & In &     \hspace{10 mm}		&  &  &  &  \\
\hline \\ [-2.5ex]
\end{tabular} 
\begin{tablenotes}
\item [a] \textit{Published near-transit radial velocities from \citet{2014arXiv1410.3449A} that have been adjusted for the velocity offset between the data sets.}.
\end{tablenotes}
\end{threeparttable}}
\end{table}

\begin{table}
\centering
\caption{Radial velocities for WASP-66 (fiber and order averaged) taken on 2014 March 21}
\label{table:WASP-66_RVs}
\centering
\resizebox{\columnwidth}{!}{%
\begin{threeparttable}[b]
\begin{tabular}{l c c c c l c c c}
\hline\hline \\ [-2.0ex]
Time & RV & S/N at & In/Out &  & Time & RV & S/N at & In/Out \\
BJD-2400000 & (\mos) & $\lambda$=5490\AA{} & Transit &  & BJD-2400000 & (\mos) & $\lambda$=5490\AA{} & Transit \\ [0.5ex]
\hline \\ [-2.0ex]
56738.02316 & -9967 $\pm$ 60 & 39 & Out &     \hspace{10 mm}		& 56738.15050 & -10088 $\pm$ 53 & 37 & In \\
56738.03908 & -9960 $\pm$ 58 & 39 & Out &     \hspace{10 mm}		& 56738.16642 & -10108 $\pm$ 61 & 38 & In \\
56738.05499 & -9973 $\pm$ 58 & 39 & In &     \hspace{10 mm}		& 56738.18233 & -10174 $\pm$ 62 & 39 & In \\
56738.07090 & -9991 $\pm$ 56 & 38 & In &     \hspace{10 mm}		& 56738.19824 & -10172 $\pm$ 53 & 37 & In \\
56738.08682 & -9982 $\pm$ 63 & 37 & In &     \hspace{10 mm}		& 56738.21416 & -10103 $\pm$ 65 & 39 & In \\
56738.10273 & -9925 $\pm$ 59 & 38 & In &     \hspace{10 mm}		& 56738.23007 & -10085 $\pm$ 56 & 39 & Out \\
56738.11865 & -10018 $\pm$ 54 & 38 & In &     \hspace{10 mm}		& 56738.24599 & -10051 $\pm$ 58 & 39 & Out \\
56738.13456 & -9988 $\pm$ 60 & 37 & In &     \hspace{10 mm}		&  &  &  &  \\
\hline \\ [-2.5ex]
\end{tabular} 
\end{threeparttable}}
\end{table}

\subsection{WASP-103 Results} \label{sec:wasp-103_results}
\begin{figure}
\centering
\includegraphics[width=1.0\linewidth]{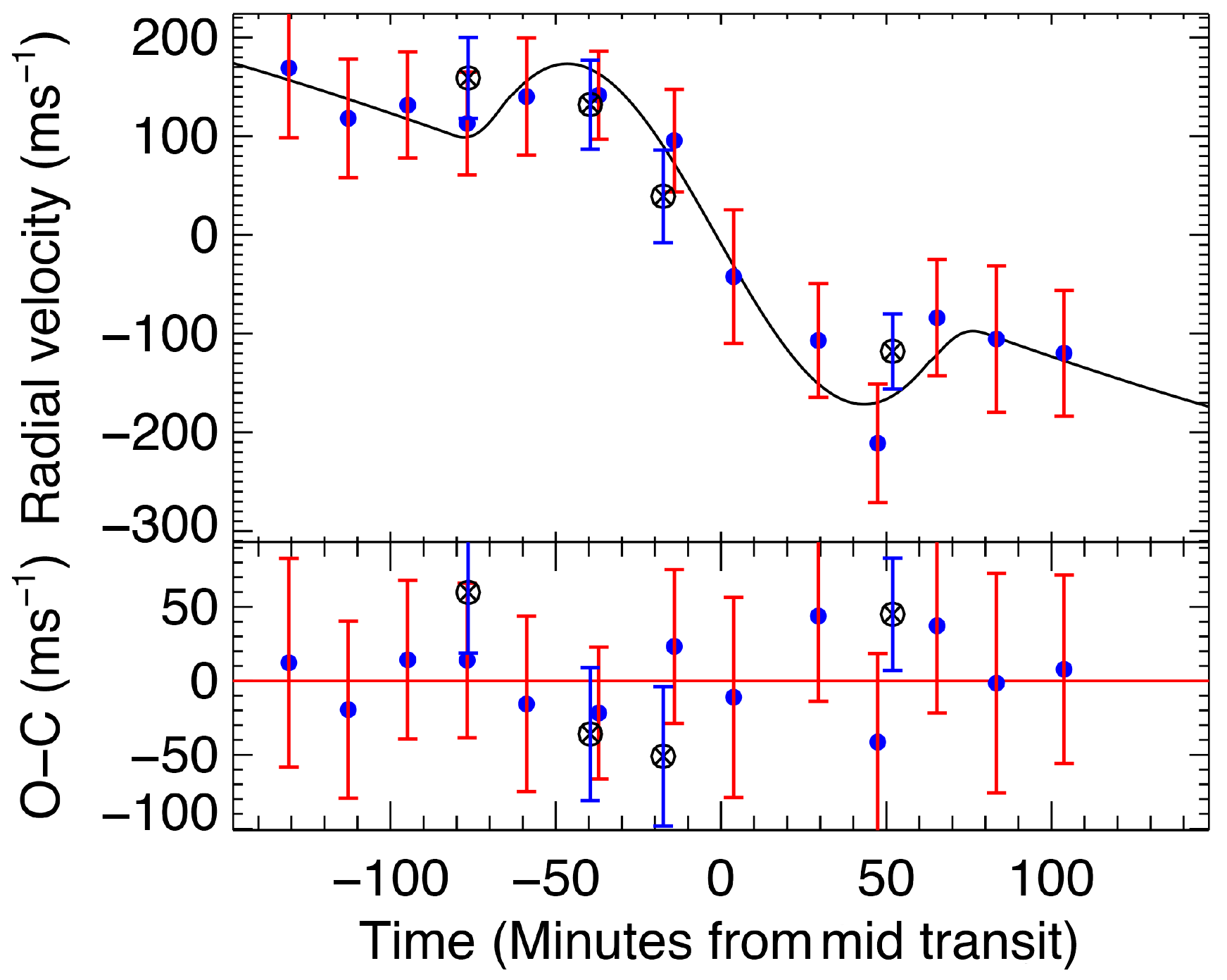}
\caption{Spectroscopic radial velocities of the WASP-103 transit. Velocities from just before, during, and after the transit are plotted as a function of time (minutes from mid-transit at 2456800.19903\,HJD) along with the best fitting model and corresponding residuals. The filled blue circles with red error bars are radial velocities obtained in this work on 2014 May 22. The four black circles with an x and with blue error bars are previously published velocities by \citet{2014A&A...562L...3G} using their quoted uncertainties. The zero velocity offset for the data set presented here was determined from the \citet{2014A&A...562L...3G} out-of-transit radial velocities.
\\ [+1.5ex]
(A color version of this figure will be available in the online journal.)}
\label{fig:WASP-103_RM}
\end{figure}

The best fit parameters and their $1\sigma$ uncertainties for WASP-103 are given in Table\,\ref{table:WASP-103_Parameters}. Figure\,\ref{fig:WASP-103_RM} shows the modeled Rossiter--McLaughlin anomaly with the observed velocities overplotted.

\begin{figure}
\centering
\includegraphics[width=1.0\linewidth]{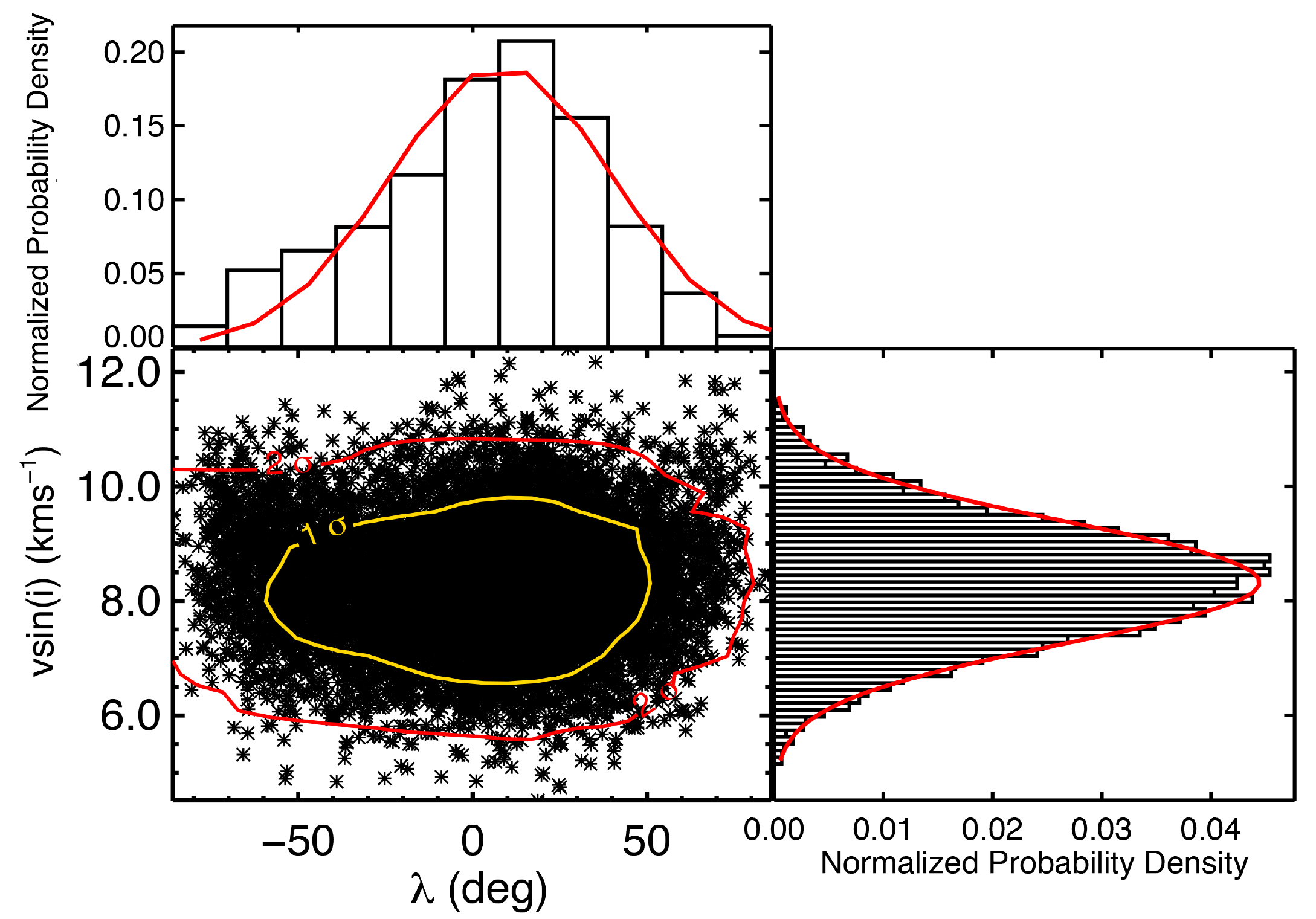}
\caption{Posterior probability distribution of $\lambda$ and $v\sin i_{\star}$ from the MCMC simulation of WASP-103. The contours show the 1 and 2 $\sigma$ confidence regions (in yellow and red, respectively). We have marginalized over $\lambda$ and $v\sin i_{\star}$ and have fit them with Gaussians (in red). This plot indicates that the distribution is somewhat non-Gaussian and suggest that there are some correlations between $\lambda$ and $v\sin i_{\star}$.
\\ [+1.5ex]
(A color version of this figure will be available in the online journal.)}
\label{fig:wasp-103_distro}
\end{figure}

The posterior probability distribution for $\lambda$ and $v\sin i_{\star}$ resulting from our MCMC simulations are shown in Figure \ref{fig:wasp-103_distro}. The $1\sigma$ and $2\sigma$ confidence contours are plotted, along with normalized density functions marginalized over $\lambda$ and $v\sin i_{\star}$ with fitted Gaussians. A non-Gaussian distribution for $\lambda$ and $v\sin i_{\star}$ is observed, suggesting that the parameters are correlated with each other or with other input parameters. $\lambda$ and $v\sin i_{\star}$ are typically degenerate when the transit impact parameter is small ($b\leq0.25$).

The Rossiter--McLaughlin effect is seen as a positive anomaly between $\sim\!80$ minutes prior to mid-transit and mid-transit and then as a negative anomaly between mid-transit and $\sim\!80$ minutes after mid-transit. This indicates that the planet first transits across the blue-shifted hemisphere during ingress and then across the red-shifted hemisphere during egress, producing a nearly symmetrical velocity anomaly as shown in Figure\,\ref{fig:WASP-103_RM}. Therefore, the orbit of WASP-103b is nearly aligned with the spin axis of its host star (i.e., that is the system is in ``spin--orbit alignment''). 

A projected obliquity of $\lambda = 3^{\circ} \pm 33^{\circ}$ and a $v\sin i_{\star}=6.5\pm2.0$\,\kms\ was obtained for this system. The $v\sin i_{\star}$ measured from the Rossiter--McLaughlin effect is anomalously low compared to the value determined from the least-squares deconvolution method ($8.8 \pm 0.7$\,\kms) and reported by \citet{2014A&A...562L...3G} as $10.6 \pm 0.9$\,\kms. Additional radial velocities covering the WASP-103b transit are needed in order to obtain a more precise spin--orbit angle.

\FloatBarrier

\subsection{WASP-87 Results} \label{sec:wasp-87_results}
\begin{figure}
\centering
\includegraphics[width=1.0\linewidth]{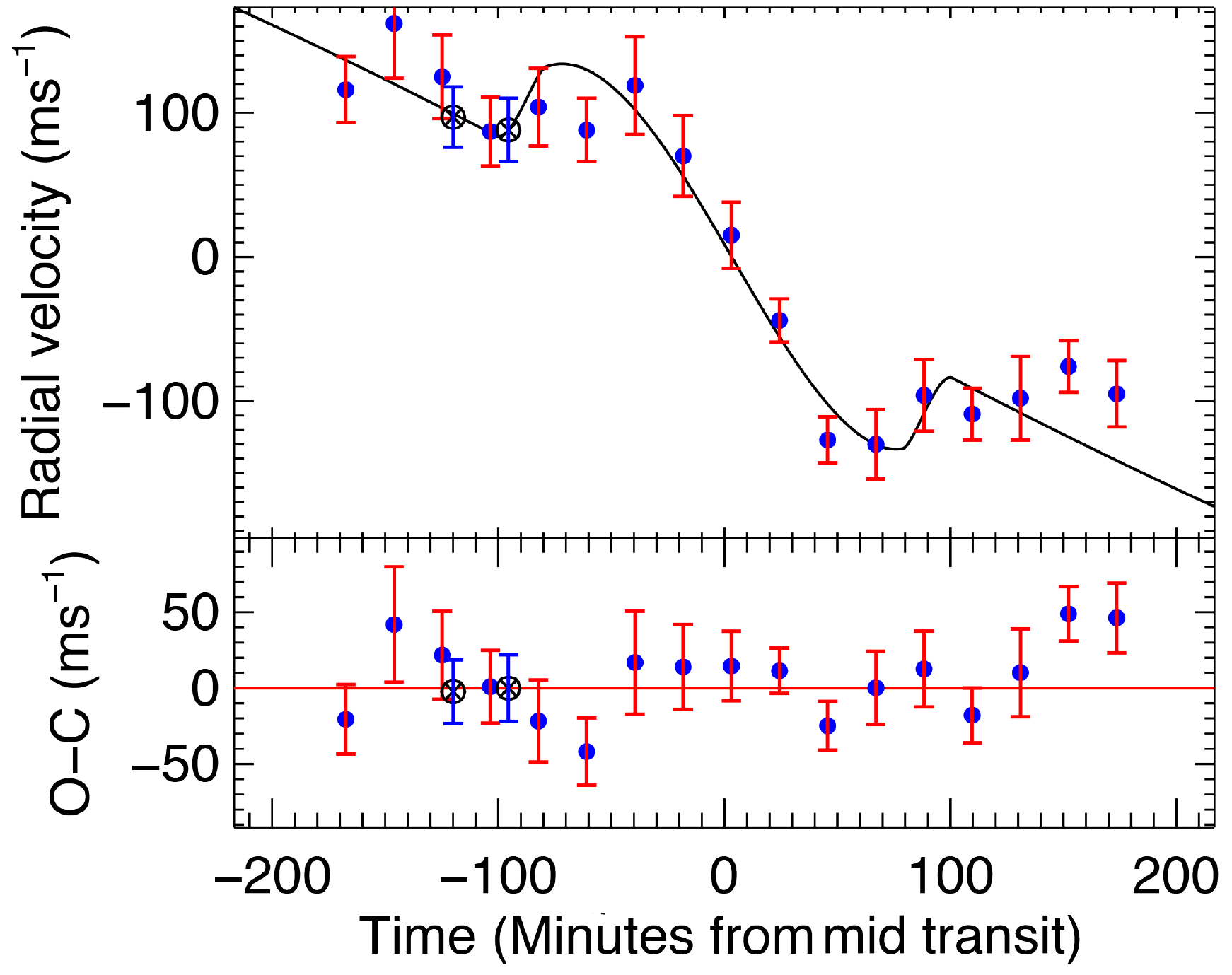}
\caption{Spectroscopic radial velocities of the WASP-87 transit. Velocities from just before, during, and after the transit are plotted as a function of time (minutes from mid-transit at 2457082.07656\,HJD) along with the best fitting model and corresponding residuals. The filled blue circles with red error bars are radial velocities obtained in this work on 2015 February 28. The two black circles with an x and with blue error bars are previously published velocities by \citet{2014arXiv1410.3449A} using their quoted uncertainties. The zero velocity offset for the data set presented here was determined from the \citet{2014arXiv1410.3449A} out-of-transit radial velocities.
\\ [+1.5ex]
(A color version of this figure will be available in the online journal.)}
\label{fig:WASP-87_RM}
\end{figure}

We determined the best fit projected obliquity and $v\sin i_{\star}$ for WASP-87b as $\lambda = -8^{\circ} \pm 11^{\circ}$ and $v\sin i_{\star}=9.8\pm0.6$\,\kms\ (using a prior on $v\sin i_{\star}$), respectively, as given in Table\,\ref{table:WASP-87_Parameters}. The $v\sin i_{\star}$ measured from the Rossiter--McLaughlin effect is in agreement with the value reported by \citet{2014arXiv1410.3449A} of $9.6 \pm 0.7$\,\kms. Figure\,\ref{fig:WASP-87_RM} shows the modeled Rossiter--McLaughlin anomaly with the observed velocities overplotted.


This system appears to be well aligned as shown by the nearly symmetrical velocity anomaly in Figure\,\ref{fig:WASP-87_RM} and moderate misalignments ($\lambda\ge22.5^{\circ}$) can be ruled out by $>2\sigma$. The Rossiter--McLaughlin effect is seen as a positive anomaly between $\sim\!100$ minutes prior to mid-transit and mid-transit and then as a negative anomaly between mid-transit and $\sim\!100$ minutes after mid-transit.

\begin{figure}
\centering
\includegraphics[width=1.0\linewidth]{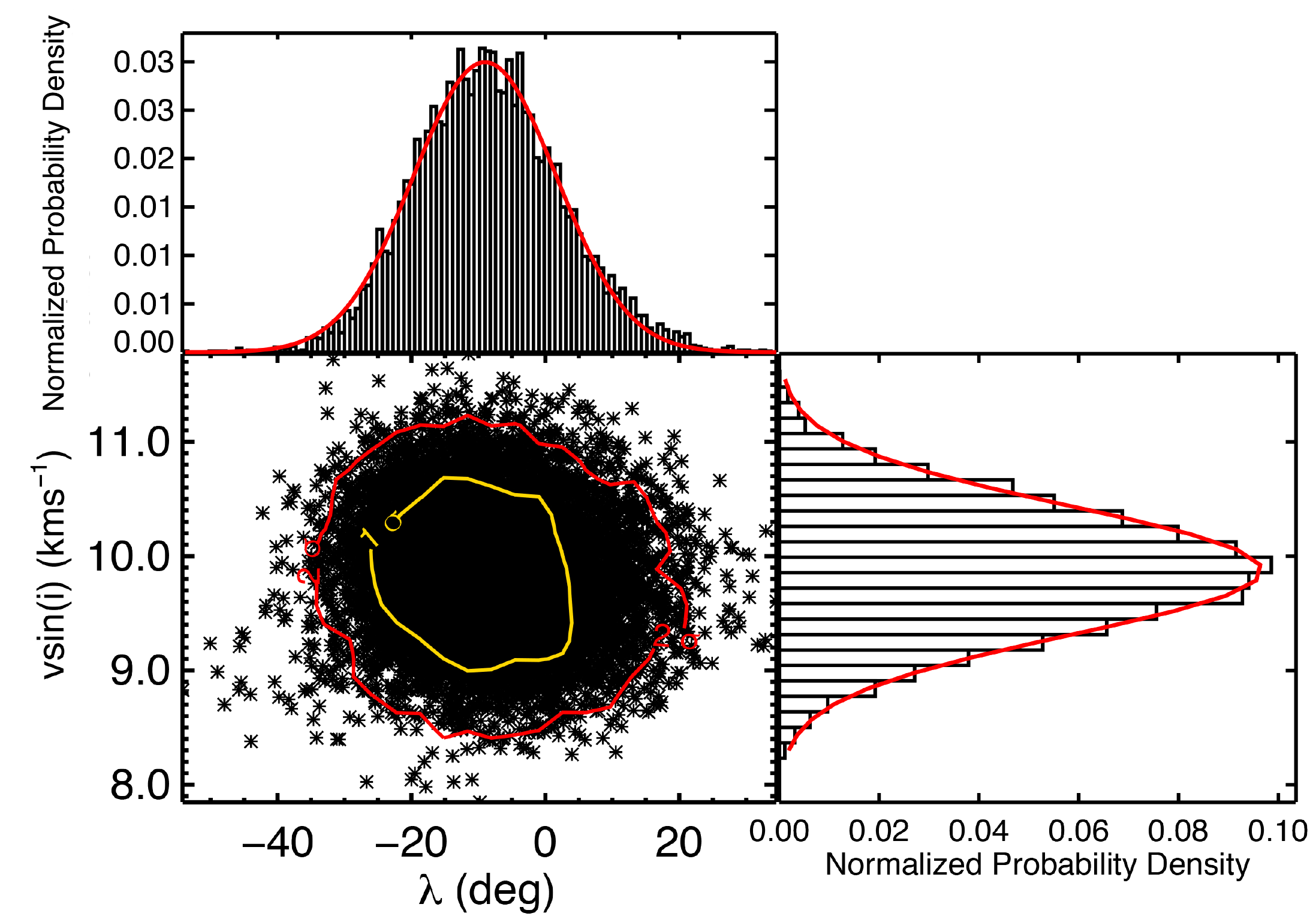}
\caption{Posterior probability distribution of $\lambda$ and $v\sin i_{\star}$ from the MCMC simulation of WASP-87. The contours show the 1 and 2 $\sigma$ confidence regions (in yellow and red, respectively). We have marginalized over $\lambda$ and $v\sin i_{\star}$ and have fit them with Gaussians (in red). The distribution appears to be Gaussian suggesting only weak correlations between $\lambda$ and $v\sin i_{\star}$.
\\ [+1.5ex]
(A color version of this figure will be available in the online journal.)}
\label{fig:wasp-87_distro}
\end{figure}

The posterior probability distribution for $\lambda$ and $v\sin i_{\star}$ resulting from our MCMC simulations are shown in Figure \ref{fig:wasp-87_distro}. The $1\sigma$ and $2\sigma$ confidence contours are plotted, along with normalized density functions marginalized over $\lambda$ and $v\sin i_{\star}$ with fitted Gaussians.

\FloatBarrier

\subsection{WASP-66 Results} \label{sec:wasp-66_results}
\begin{figure}
\centering
\includegraphics[width=1.0\linewidth]{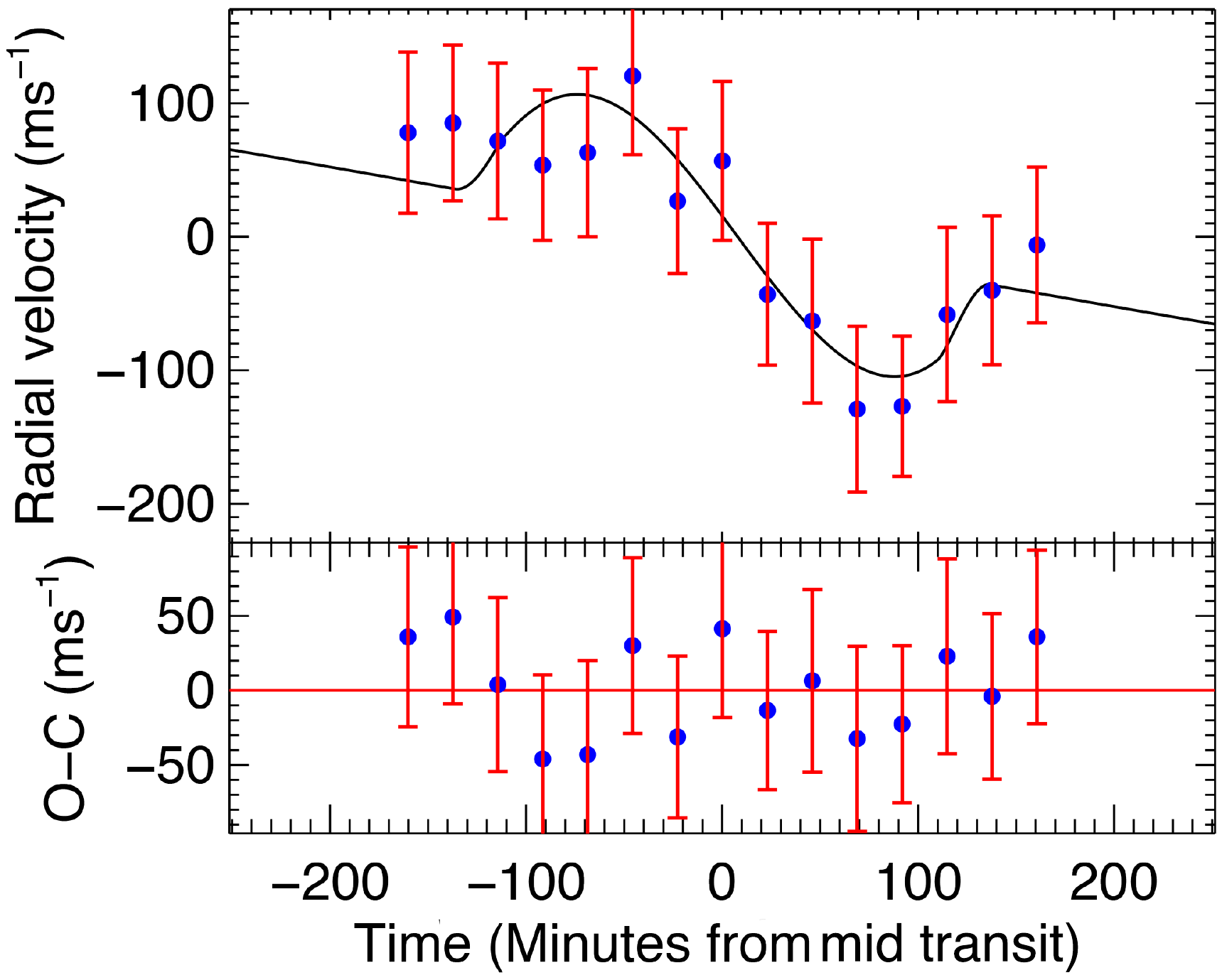}
\caption{Spectroscopic radial velocities of the WASP-66 taken just before, during, and after the transit. These are plotted as a function of time (minutes from mid-transit at 2456738.13445\,HJD) along with the best fitting model and corresponding residuals. The filled blue circles with red error bars are radial velocities obtained in this work on 2014 March 21. The zero velocity offset for our data set was determined from the \citet{2012MNRAS.426..739H} out-of-transit radial velocities (not shown). The velocities appear to be anomalously below the best fit model of the Rossiter--McLaughlin effect during the first half of the transit. The cause for this is unknown but might be due to the planet transiting over a star spot, random noise in the data, or systematic effects that have not been accounted for in producing the radial velocities.
\\ [+1.5ex]
(A color version of this figure will be available in the online journal.)}
\label{fig:WASP-66_RM_fixed}
\end{figure}

Table\,\ref{table:WASP-66_Parameters} presents the best fit parameters and their $1\sigma$ uncertainties for WASP-66. The best fit solution for $\lambda$ is $\lambda = -4^{\circ} \pm 22^{\circ}$ and $v\sin i_{\star} = 12.1 \pm 2.2$\,\kms. The $v\sin i_{\star}$ measured from the Rossiter--McLaughlin effect is consistent (to within the measured uncertainties) with the value reported by \citet{2012MNRAS.426..739H} of $13.4 \pm 0.9$\,\kms\ and the value we determined using the least-squares deconvolution method ($11.8 \pm 0.4$\,\kms). High obliquity orbits can be ruled from our data, but additional radial velocities covering the WASP-66b transit are needed to lock down a more precise spin--orbit angle. A modeled Rossiter--McLaughlin anomaly with the observed velocities overplotted is shown in Figure\,\ref{fig:WASP-66_RM_fixed}.



\begin{figure}
\centering
\includegraphics[width=1.0\linewidth]{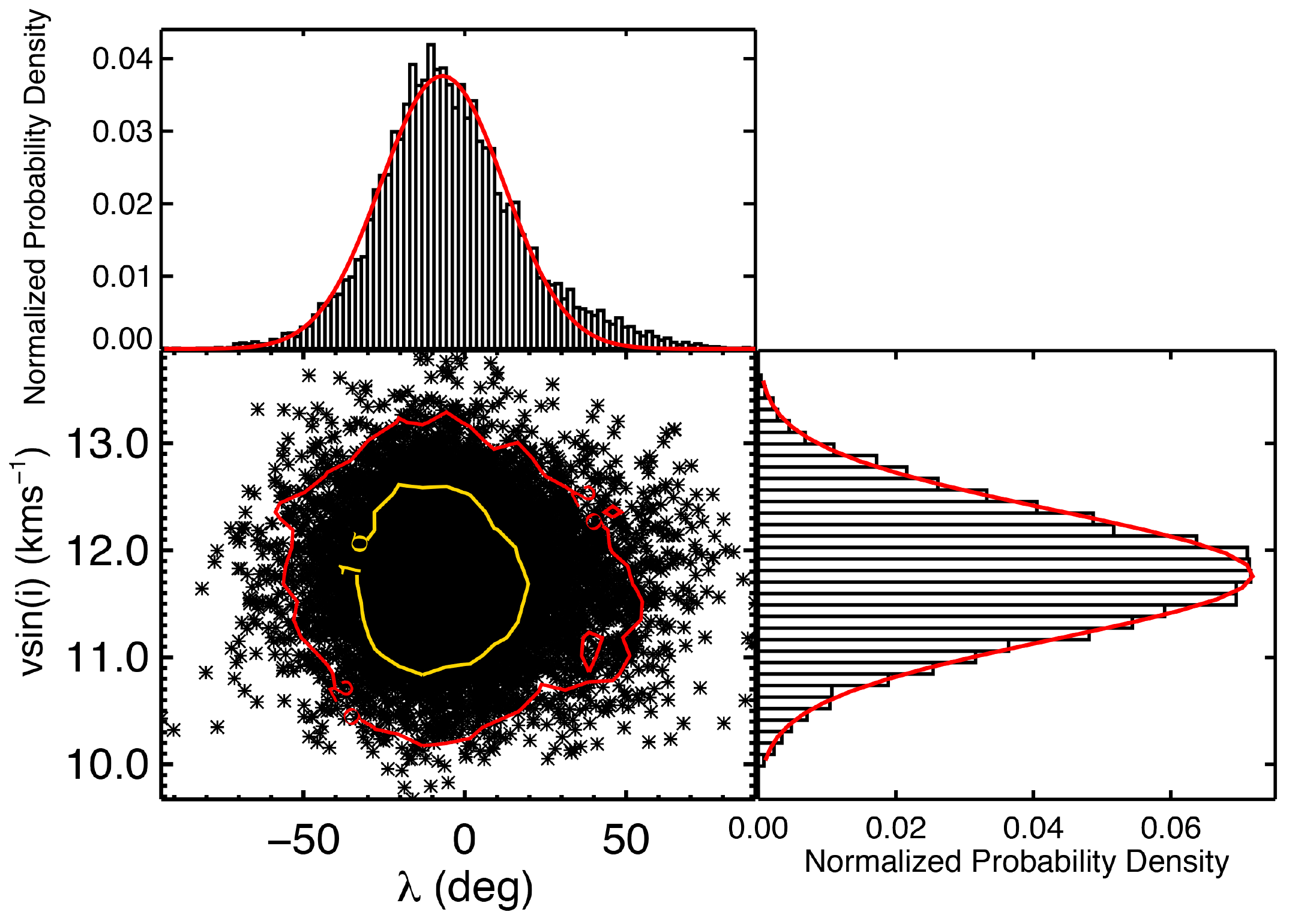}
\caption{Posterior probability distribution of $\lambda$ and $v\sin i_{\star}$ from the MCMC simulation of WASP-66. The contours show the 1 and 2 $\sigma$ confidence regions (in yellow and red, respectively). Fitted Gaussians (in red) are shown separately for the marginalized distributions over $\lambda$ (above) and $v\sin i_{\star}$ (right).
\\ [+1.5ex]
(A color version of this figure will be available in the online journal.)}
\label{fig:wasp-66_distro}
\end{figure}

Figure\,\ref{fig:wasp-66_distro} shows the resulting posterior probability distribution of $\lambda$ and $v\sin i_{\star}$, including the locations of the $1\sigma$ and $2\sigma$ confidence contours, for WASP-66 from the MCMC simulations. We have also produced normalized density functions, marginalized over $\lambda$ and $v\sin i_{\star}$ and with fitted Gaussians, as shown in Figure\,\ref{fig:wasp-66_distro}. The resulting distribution for $\lambda$ and $v\sin i_{\star}$ appears to be Gaussian, indicating that these parameters are mostly uncorrelated.


Similar to WASP-103b and WASP-87b, the observed Rossiter--McLaughlin effect for WASP-66b is seen as a positive anomaly between $\sim\!140$ minutes prior to mid-transit and mid-transit and then as a negative anomaly between mid-transit and $\sim\!140$ minutes after mid-transit, as shown in Figure\,\ref{fig:WASP-66_RM_fixed}. This indicates that the orbit of WASP-66b is nearly aligned with the spin axis of its host star.

\begin{table*}
\centering
\begin{threeparttable}[b]
\caption{System parameters for WASP-103}
\centering
\begin{tabular}{l c}
\hline\hline \\ [-2.0ex]
Parameter & Value \\ [0.5ex]
\hline\hline \\ [-2.0ex]
{\textit{Parameters as given by \citet{2014A&A...562L...3G}}} \\ [0.5ex]
{\textit{and used as priors in model}} \\
\hline \\ [-2.0ex]
Mid-transit epoch (2400000-HJD)\tnote{a}, $T_{0}$ & $56800.19903 \pm 0.00075$ \\
Orbital period\tnote{a}, $P$ & $0.925542 \pm 0.000019$\,days \\
Semimajor axis\tnote{a}, $a$ & $0.01985 \pm 0.00021$\,AU \\
Orbital inclination\tnote{a}, $I$ & $86.3^{\circ} \pm 2.7^{\circ}$ \\
Impact parameter\tnote{a}, $b$ & $0.19 \pm 0.13$ \\
Transit depth\tnote{a}, $(R_{P}/R_{\star})^{2}$ & $1.195^{+0.042}_{-0.038}$ \\
Orbital eccentricity\tnote{b}, $e$ & 0.0 (assumed) \\
Argument of periastron\tnote{b}, $\varpi$ & N/A ($e=0$) \\
Stellar reflex velocity\tnote{b}, $K_{\star}$ & $271 \pm 15$\,\mos \\
Stellar mass\tnote{a}, $M_{\star}$ & $1.220^{+0.039}_{-0.036}$\,$M_{\odot}$ \\
Stellar radius\tnote{a}, $R_{\star}$ & $1.436^{+0.052}_{-0.031}$ $R_{\odot}$ \\
Planet mass\tnote{a}, $M_{P}$ & $1.490 \pm 0.088$\,$M_{J}$ \\
Planet radius\tnote{a}, $R_{P}$ & $1.528^{+0.073}_{-0.047}$ $R_{J}$ \\
Stellar micro-turbulence\tnote{b}, $\xi_{t}$ & $1.1 \pm 0.2$\,\kms \\
Stellar macro-turbulence\tnote{b}, $v_\mathrm{mac}$ & N/A  \\
Stellar limb-darkening coefficient\tnote{c}, $q_{1}$ & 0.3999 (adopted) \\
Stellar limb-darkening coefficient\tnote{c}, $q_{2}$ & 0.2939 (adopted) \\
Stellar effective temperature\tnote{d}, $T_\mathrm{eff}$ & $6110 \pm 160$\,K (adopted) \\
Velocity at published epoch $T_{P}$\tnote{b}, $V_{T_{P}}$ & $-42.001 \pm 0.005$\,\kms \\
RV offset between Gillon et al. and AAT data set\tnote{a}, $V_{d}$ & $32 \pm 29$\,\mos \\ 
[0.5ex]
\hline\hline \\ [-2.0ex]
\textit{Parameters determined from a Markov Chain Monte Carlo model fit} \\ [0.5ex]
\textit{using AAT velocities.} \\ [0.5ex]
\hline \\ [-2.0ex]
Projected obliquity angle\tnote{e}, $\lambda$ & $3^{\circ} \pm 33^{\circ}$ \\ [0.5ex]
Projected stellar rotation velocity\tnote{f}, ${v\sin i_{\star}}$ & $6.5 \pm 2.0$\,\kms \\ [0.5ex]
\hline\hline \\ [-2.0ex]
\textit{Independent measurement of $v\sin i_{\star(Ind)}$ from LSD} \\ [0.5ex]
{\textit{method and \citet{2014A&A...562L...3G} $v\sin i_{\star(G)}$ published value.}} \\ [0.5ex]
\hline \\ [-2.0ex]
Projected stellar rotation velocity, $v\sin i_{\star(Ind)}$ & $8.8 \pm 0.7$\,\kms \\ [0.5ex]
Projected stellar rotation velocity, $v\sin i_{\star(G)}$ & $10.6 \pm 0.9$\,\kms \\ [0.5ex]
\hline 
\end{tabular}
\vspace{1mm}
\label{table:WASP-103_Parameters}
\begin{tablenotes}
\item [a] \textit{Prior parameters fixed to the indicated value for final fit, but allowed to vary in the MCMC for uncertainty estimation.}
\item [b] \textit{Parameters fixed at values given by \cite{2014A&A...562L...3G}.}
\item [c] \textit{Limb darkening coefficients determined from look up tables in \cite{2011A&A...529A..75C}.}
\item [d] \textit{Effective temperature from \citet{2014A&A...562L...3G} and used to determine limb-darkening coefficients.}
\item [e] \textit{$\lambda$ obtained by imposing a prior on ${v\sin i_{\star}}$ of $v\sin i_{\star(Ind)}$ and $2\sigma_{v\sin i_{\star(Ind)}}$.}
\item [f] \textit{No informative prior was imposed on ${v\sin i_{\star}}$ to obtain the best fit value and uncertainty.}
\end{tablenotes}
\end{threeparttable} 
\end{table*}

\begin{table*}
\centering
\begin{threeparttable}[b]
\caption{System parameters for WASP-87}
\centering
\begin{tabular}{l c}
\hline\hline \\ [-2.0ex]
Parameter & Value \\ [0.5ex]
\hline\hline \\ [-2.0ex]
{\textit{Parameters as given by \citet{2014arXiv1410.3449A}}} \\ [0.5ex]
{\textit{and used as priors in model}} \\
\hline \\ [-2.0ex]
Mid-transit epoch (2400000-HJD)\tnote{a}, $T_{0}$ & $57082.07656 \pm 0.00021$ \\
Orbital period\tnote{a}, $P$ & $1.6827950 \pm 0.0000019$\,days \\
Semimajor axis\tnote{a}, $a$ & $0.02946 \pm 0.00075$\,AU \\
Orbital inclination\tnote{a}, $I$ & $81.07^{\circ} \pm 0.63^{\circ}$ \\
Impact parameter\tnote{a}, $b$ & $0.604 \pm 0.028$ \\
Transit depth\tnote{a}, $(R_{P}/R_{\star})^{2}$ & $0.00765 \pm 0.00013$ \\
Orbital eccentricity\tnote{b}, $e$ & 0.0 (assumed) \\
Argument of periastron\tnote{c}, $\varpi$ & N/A ($e=0$) \\
Stellar reflex velocity\tnote{c}, $K_{\star}$ & $325 \pm 14$\,\mos \\
Stellar mass\tnote{c}, $M_{\star}$ & $1.204 \pm 0.093$\,$M_{\odot}$ \\
Stellar radius\tnote{a}, $R_{\star}$ & $1.627 \pm 0.062$ $R_{\odot}$ \\
Planet mass\tnote{c}, $M_{P}$ & $2.18 \pm 0.15$\,$M_{J}$ \\
Planet radius\tnote{a}, $R_{P}$ & $1.385 \pm 0.060$ $R_{J}$ \\
Stellar micro-turbulence\tnote{c}, $\xi_{t}$ & $1.34 \pm 0.13$\,\kms \\
Stellar macro-turbulence\tnote{c}, $v_\mathrm{mac}$ & $5.9 \pm 0.6$  \\
Stellar limb-darkening coefficient\tnote{d}, $q_{1}$ & 0.3749 (adopted) \\
Stellar limb-darkening coefficient\tnote{d}, $q_{2}$ & 0.2669 (adopted) \\
Stellar effective temperature\tnote{e}, $T_\mathrm{eff}$ & $6450 \pm 120$\,K (adopted) \\
Velocity at published epoch $T_{P}$\tnote{c}, $V_{T_{P}}$ & $-14.1845 \pm 0.0079$\,\kms \\
RV offset between Anderson et al. and AAT data set\tnote{a}, $V_{d}$ & $770 \pm 12$\,\mos \\
[0.5ex]
\hline\hline \\ [-2.0ex]
\textit{Parameters determined from a Markov Chain Monte Carlo model fit} \\ [0.5ex]
\textit{using AAT velocities.} \\ [0.5ex]
\hline \\ [-2.0ex]
Projected obliquity angle\tnote{f}, $\lambda$ & $-8^{\circ} \pm 11^{\circ}$ \\ [0.5ex]
Projected stellar rotation velocity\tnote{g}, $v\sin i_{\star}$ & $9.9 \pm 0.6$\,\kms \\ [0.5ex]
\hline\hline \\ [-2.0ex]
{\textit{\citet{2014arXiv1410.3449A} $v\sin i_{\star(A)}$ published value.}} \\ [0.5ex]
\hline \\ [-2.0ex]
Projected stellar rotation velocity, $v\sin i_{\star(A)}$ & $12.2 \pm 1.9$\,\kms \\ [0.5ex]
\hline 
\end{tabular}
\vspace{1mm}
\label{table:WASP-87_Parameters}
\begin{tablenotes}
\item [a] \textit{Prior parameters fixed to the indicated value for final fit, but allowed to vary in the MCMC for uncertainty estimation.}
\item [b] \textit{Parameter fixed to zero.}
\item [c] \textit{Parameters fixed at values given by \citet{2014arXiv1410.3449A}.}
\item [d] \textit{Limb-darkening coefficients determined from look up tables in \citet{2011A&A...529A..75C}.}
\item [e] \textit{Effective temperature from \citet{2014arXiv1410.3449A} and used to determine limb darkening coefficients.}
\item [f] \textit{$\lambda$ obtained by imposing a prior on ${v\sin i_{\star}}$ of $v\sin i_{\star(Ind)}$ and $2\sigma_{v\sin i_{\star(Ind)}}$.}
\item [g] \textit{No informative prior was imposed on ${v\sin i_{\star}}$ to obtain the best fit value and uncertainty.}
\end{tablenotes}
\end{threeparttable} 
\end{table*}

\hspace*{-10cm}
\begin{table*}
\centering
\begin{threeparttable}[b]
\caption{System parameters for WASP-66}
\centering
\begin{tabular}{l c}
\hline\hline \\ [-2.0ex]
Parameter & Value \\ [0.5ex]
\hline\hline \\ [-2.0ex]
{\textit{Parameters as given by \citet{2012MNRAS.426..739H}}} \\ [0.5ex]
{\textit{and used as priors in model}} \\
\hline \\ [-2.0ex]
Mid-transit epoch (2400000-HJD)\tnote{a}, $T_{0}$ & $56738.13445 \pm 0.00035$ \\
Orbital period\tnote{a}, $P$ & $4.086052 \pm 0.000007$\,days \\
Semimajor axis\tnote{a}, $a$ & $0.0546 \pm 0.0009$\,AU \\
Orbital inclination\tnote{a}, $I$ & $85.9^{\circ} \pm 0.9^{\circ}$ \\
Impact parameter\tnote{a}, $b$ & $0.48^{+0.06}_{-0.08}$ \\
Transit depth\tnote{a}, $(R_{P}/R_{\star})^{2}$ & $0.00668 \pm 0.00016$ \\
Orbital eccentricity\tnote{b}, $e$ & 0.0 (assumed) \\
Argument of periastron\tnote{b}, $\varpi$ & N/A ($e=0$) \\
Stellar reflex velocity\tnote{b}, $K_{\star}$ & $246 \pm 11$\,\mos \\
Stellar mass\tnote{b}, $M_{\star}$ & $1.30 \pm 0.07$\,$M_{\odot}$ \\
Stellar radius\tnote{a}, $R_{\star}$ & $1.75 \pm 0.09$ $R_{\odot}$ \\
Planet mass\tnote{b}, $M_{P}$ & $2.32 \pm 0.13$\,$M_{J}$ \\
Planet radius\tnote{a}, $R_{P}$ & $1.39 \pm 0.09$ $R_{J}$ \\
Stellar micro-turbulence\tnote{b}, $\xi_{t}$ & $2.2 \pm 0.3$\,\kms \\
Stellar macro-turbulence\tnote{b}, $v_\mathrm{mac}$ & N/A  \\
Stellar limb-darkening coefficient\tnote{c}, $q_{1}$ & 0.3932 (adopted) \\
Stellar limb-darkening coefficient\tnote{c}, $q_{2}$ & 0.2619 (adopted) \\
Stellar effective temperature\tnote{d}, $T_\mathrm{eff}$ & $6600 \pm 150$\,K (adopted) \\
Velocity at published epoch $T_{P}$\tnote{b}, $V_{T_{P}}$ & $-10.02458 \pm 0.00013$\,\kms \\
RV offset between Hellier et al. and AAT data set\tnote{a}, $V_{d}$ & $-33^{+22}_{-23}$\,\mos \\ [0.5ex]
\hline\hline \\ [-2.0ex]
\textit{Parameters determined from a Markov Chain Monte Carlo model fit} \\ [0.5ex]
\textit{using AAT velocities.} \\ [0.5ex]
\hline \\ [-2.0ex]
Projected obliquity angle\tnote{e}, $\lambda$ & $-4^{\circ} \pm 22^{\circ}$ \\ [0.5ex]
Projected stellar rotation velocity\tnote{f}, $v\sin i_{\star}$ & $12.1 \pm 2.2$\,\kms \\ [0.5ex]
\hline\hline \\ [-2.0ex]
\textit{Independent measurement of $v\sin i_{\star(Ind)}$ from LSD} \\ [0.5ex]
{\textit{method and \citet{2012MNRAS.426..739H} $v\sin i_{\star(H)}$ published value.}} \\ [0.5ex]
\hline \\ [-2.0ex]
Projected stellar rotation velocity, $v\sin i_{\star(Ind)}$ & $11.8 \pm 0.4$\,\kms \\ [0.5ex]
Projected stellar rotation velocity, $v\sin i_{\star(H)}$ & $13.4 \pm 0.9$\,\kms \\ [0.5ex]
\hline 
\end{tabular}
\vspace{1mm}
\label{table:WASP-66_Parameters}
\begin{tablenotes}
\item [a] \textit{Prior parameters fixed to the indicated value for final fit, but allowed to vary in the MCMC for uncertainty estimation.}
\item [b] \textit{Parameters fixed at values given by \cite{2012MNRAS.426..739H}.}
\item [c] \textit{Limb darkening coefficients determined from look up tables in \cite{2013A&A...552A..16C}.}
\item [d] \textit{Effective temperature from \citet{2012MNRAS.426..739H} and used to determine limb-darkening coefficients.}
\item [e] \textit{$\lambda$ obtained by imposing a prior on ${v\sin i_{\star}}$ of $v\sin i_{\star(Ind)}$ and $2\sigma_{v\sin i_{\star(Ind)}}$.}
\item [f] \textit{No informative prior was imposed on ${v\sin i_{\star}}$ to obtain the best fit value and uncertainty.}
\end{tablenotes}
\end{threeparttable} 
\end{table*}

\section{Discussion}\label{sec:Discussion_wasp-66-103}
We have carried out measurements of the spin--orbit alignments for three Hot Jupiters, WASP-103b, WASP-87b, and WASP-66b. Our results indicate that the three planets are in nearly aligned orbits with respect to the projected rotational axis of their host star. The spin--orbit angle measured for WASP-103b is $\lambda=3^{\circ} \pm 33^{\circ}$. The best fit $v\sin i_{\star}$ for WASP-103b using the Rossiter--McLaughlin effect and not imposing an informative prior on $v\sin i_{\star}$ is $v\sin i_{\star}=6.5 \pm 2.0$\,\kms, which is in disagreement (within uncertainties) with both the value determined from the least-squares deconvolution method and with the value of $10.6 \pm 0.9$\,\kms\ reported by \citet{2014A&A...562L...3G}. This is concluded to potentially be due stellar line broadening mechanisms other than rotation.

We measured a projected obliquity of $\lambda=-8^{\circ} \pm 11^{\circ}$ for WASP-87b. The stellar rotational velocity of WASP-87 from the Rossiter--McLaughlin effect is $v\sin i_{\star}=9.8\pm0.6$\,\kms. This is in agreement (within uncertainties) with the value of $9.6 \pm 0.7$\,\kms\ reported by \citet{2014arXiv1410.3449A}. Out of the three systems analyzed in this work, WASP-87b has the best constrained spin--orbit angle.

For WASP-66b, the spin--orbit angle is determined to be $\lambda=-4^{\circ} \pm 22^{\circ}$. The best fit $v\sin i_{\star}$ obtained using the Rossiter--McLaughlin effect is $v\sin i_{\star}=12.1 \pm 2.2$\,\kms, which is in agreement with both the value determined from least-squares deconvolution and the value of $13.4 \pm 0.9$\,\kms\ reported by \citet{2012MNRAS.426..739H}. Additional radial velocities of the Rossiter--McLaughlin effect for both WASP-103 and WASP-66 are required to rule out moderately misaligned orbits, however, polar and retrograde orbits are disfavored for both systems.

The tidal dissipation timescale was determined for WASP-103b, WASP-87b, and WASP-66b following the procedures given by \citet{2012ApJ...757...18A}. For WASP-103b, the tidal dissipation timescale is $\tau_{CE}=1.24\times10^{9}$\,years (using the convective timescale for alignment). It is helpful to normalize $\tau_{CE}$ to a useful stellar timescale of 5\,Gyr which results in $\frac{\tau_{CE}}{5\mathrm{Gyr}}=0.248$. If the mass of the convective envelope is taken into account \citep[the second approach of][]{2012ApJ...757...18A}, the timescale becomes $\tau_{mcz}=3.11\times10^{6}$\,years. Normalizing this to the model age for the host star ($T_{zams}=4.0\times10^{9}$\,years) results in $\frac{\tau_{mcz}}{T_{zams}}=7.77\times10^{-4}$.

The tidal dissipation timescale for WASP-87b is $\tau_{RA}=1.05\times10^{13}$\,years (using the radiative timescale for alignment) and $\frac{\tau_{RA}}{5\mathrm{Gyr}}=2.11\times10^{3}$. If the mass of the convective envelope is taken into consideration, then the timescale becomes $\tau_{mcz}=1.94\times10^{11}$\,years and $\frac{\tau_{mcz}}{T_{zams}}=51$ when $T_{zams}=3.8\times10^{9}$\,years is set to the model age of WASP-87.

For WASP-66b, the tidal dissipation timescale is $\tau_{RA}=1.11\times10^{15}$\,years (using the radiative timescale for alignment) and $\frac{\tau_{RA}}{5\mathrm{Gyr}}=2.22\times10^{5}$. Taking the mass of the convective envelope into consideration, the timescale becomes $\tau_{mcz}=2.41\times10^{12}$\,years and $\frac{\tau_{mcz}}{T_{zams}}=731$ when $T_{zams}=3.3\times10^{9}$\,years is set to the model age of WASP-66.

The low-obliquity orbit found for WASP-103b is expected given the very short realignment timescale for this system. The planet is likely experiencing very strong tidal forces and could be in the process of being tidally disrupted and consumed by its host star. Investigations of the orbital migration histories of other ultra-short-period Hot Jupiters through spin--orbit alignment measurements may provide clues on the processes involved in their formation and migration as well as tidal interactions they might be experiencing with their host star.

WASP-87 is a hot ($6450\pm110$\,K), metal-poor mid-F star with a relatively long radiative realignment timescale that would not be very efficient at realigning a high-obliquity planetary orbit. In addition, \citet{2014arXiv1410.3449A} found a nearby early-G star $8.2^{\texttt{"}}$ from WASP-87A that appears to be a bound companion, suggesting that Kozai resonances might have influenced the obliquity of WASP-87b. Given these circumstances, WASP-87b was predicted to have high probability of being misaligned, but was observed to be on a low-obliquity orbit.

Similarly, WASP-66 is a hot ($T_{eff}=6600\pm150$\,K), mid-F primary with a long radiative realignment timescale and was also predicted to have high probability of being misaligned. However, the orbital obliquity for this planet is low (though moderate misalignments cannot be ruled out). Orbital obliquities are thought to be distributed randomly from the migration processes that produce Hot Jupiters, regardless of the value of $T_\mathrm{eff}$ \citep{2010ApJ...718L.145W,2012ApJ...757...18A}. Therefore, the observed obliquities should be randomly distributed for systems with long $\tau_{mcz}$ and low obliquites for systems with short $\tau_{mcz}$ \citep{2010ApJ...718L.145W,2012ApJ...757...18A}. Alternatively, WASP-66b and WASP-87b could have undergone type 1 and 2 disk-driven migration \citep[e.g.,][]{1996Natur.380..606L} and therefore never had their orbits misaligned by the migration process.

It is important to emphasize that the true spin--orbit angle ($\psi$) cannot be determined directly by the Rossiter--McLaughlin effect \citep[e.g., see][]{2009ApJ...696.1230F}. Instead, we measure the sky-projected spin--orbit angle ($\lambda$), since the orientation of the stellar rotation axis to our line of sight is unknown. \citet{2009ApJ...696.1230F}, \citet{2011ApJ...729..138M}, and \citet{2013ApJ...766..101C} have shown that $\lambda$ is only a lower limit on $\psi$ and that a small $\left| \lambda \right|$ does not necessarily translate as a small true misalignment ($\left| \psi \right|$). However, a large $\left| \lambda \right|$ does indicate a large value for $\left| \psi \right|$ \citep{2009ApJ...696.1230F}. There are ways of constraining the true spin--orbit angle \citep[see, e.g.,][]{2009ApJ...696.1230F,2011ApJ...729..138M,2011exha.book.....P,2013ApJ...766..101C}. One way to estimate the inclination of a star's spin axis ($i_{\star}$) is by combining the projected rotational velocity ($v\sin i$ from line broadening), the stellar rotational period $P_{rot}$ (from high precision photometry), and $R_{\star}$ (from accurate knowledge of the spectral type). Then using $\sin i_{\star}$, $I$, and $\lambda$, one can determine $\psi$ \citep[see][]{2013ApJ...766..101C}. Asteroseismology is a powerful technique that provides an additional means of estimating $\sin i_{\star}$ and determining $\psi$ for transiting planets, independent of the Rossiter--McLaughlin effect \citep[e.g., see][]{2013ApJ...766..101C}.

\citet{2010ApJ...718L.145W} and \citet{2012ApJ...757...18A} noted a dependence between the effective temperature of a host star and the degree of orbital misalignment of its planet. They observed that obliquities generally fall into two distinct populations. The coolest stars ($T < 6250$\,K) tend to host planets with well aligned orbits while stars hotter than $6250$\,K host planets with a wide distribution of orbital obliquities. 

\begin{figure}
\centering
\includegraphics[width=1.0\linewidth]{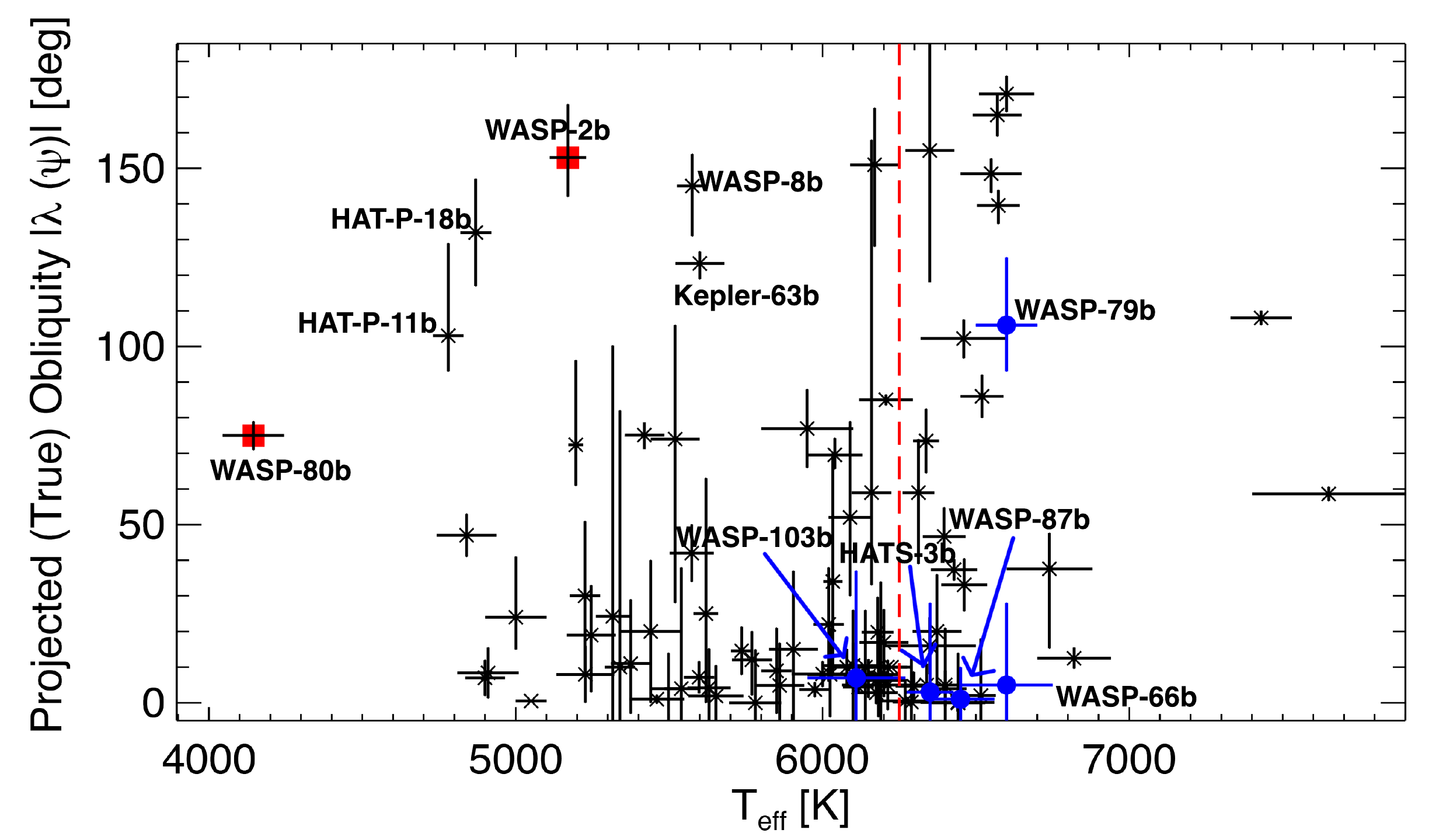}
\caption{Projected orbital obliquity ($\lambda$) of exoplanets as a function of their host star's stellar effective temperature ($T_\mathrm{eff}$). We have updated this figure from \citet{2014ApJ...792..112A} to include WASP-103b, WASP-87b, and WASP-66b as measured here as well as the systems with newly published spin--orbit angles. The red dashed line indicates the $T_\mathrm{eff}=6250$\,K boundary between thin and thick stellar convective zones that influence the strength of planet--star tidal interactions that dissipate orbital obliquities. The filled blue circles represent the systems WASP-103b, WASP-87b, WASP-66b, HATS-3b \citep{2014ApJ...792..112A}, and WASP-79b \citep{2013ApJ...774L...9A}. The systems labeled to the left of the red dashed line have anomalously large obliquities that break the observed trend of cool stars ($T_{eff}<6250$\,K) hosting planets on low-obliquity orbits (WASP-80b, \citealt{2013A&A...551A..80T}; HAT-P-11b, \citealt{2010ApJ...723L.223W}; HAT-P-18b, \citealt{2014A&A...564L..13E}; WASP-2b, \citealt{2010A&A...524A..25T}; Kepler-63b, \citealt{2013ApJ...775...54S}; and WASP-8b, \citealt{2010A&A...517L...1Q}). The $\lambda$ values are not well constrained for the two planets (WASP-80b, \citealt{2013A&A...551A..80T}; and WASP-2b, \citealt{2011ApJ...738...50A}) marked in red squares.
\\ [+1.5ex]
(A color version of this figure will be available in the online journal.)}
\label{Figure:obliquity_temp}
\end{figure}

Figure\,\ref{Figure:obliquity_temp} shows an updated version of the projected orbital obliquity verses stellar temperature plot from \citet{2014A&A...564L..13E} and \citet{2014ApJ...792..112A}, with our measured spin--orbit angles for WASP-103b, WASP-87b, and WASP-66b and from our recently published work on WASP-79b \citep{2013ApJ...774L...9A} and HATS-3b \citep{2014ApJ...792..112A}. This figure suggests that the temperature trend observed in previous studies is no longer so clear cut. A substantial number of stars with $T_\mathrm{eff}<6250$\,K host planets on misaligned orbits. Of the 68 stars with $T_\mathrm{eff}<6250$\,K that have measured spin--orbit angles, 21 ($\sim\!31\%$) host planets on misaligned orbits.

\begin{figure}
\centering
\includegraphics[width=1.0\linewidth]{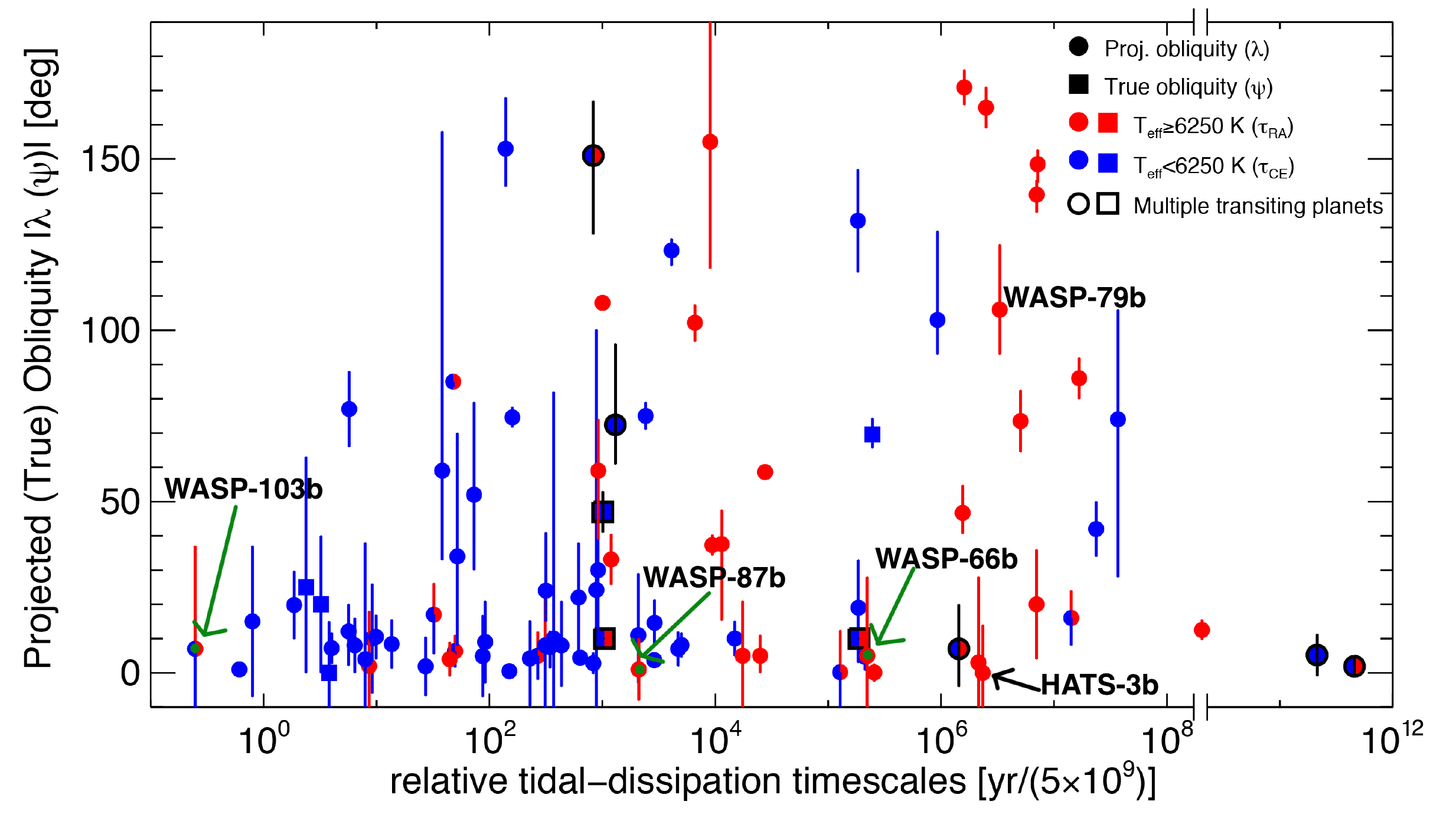}
\caption{Projected spin--orbit alignments of exoplanetary systems as a function of their relative alignment timescale for stars with either convective (CE) or radiative envelopes (RA), calibrated from binary studies. We have updated this figure from \citet{2013ApJ...771...11A} and \citet{2014ApJ...792..112A} to include the obliquity measurements of WASP-103b, WASP-87b, and WASP-66b (indicated by the arrow and green dot), as well as systems with spin--orbit angles measured in the literature since these publications. Stars that have effective temperatures higher than $6250$\,K are represented by filled red circles and squares with red error bars while blue filled circles and squares with blue error bars are for stars with effective temperatures less than $6250$\,K. The circles and squares that are half red and blue show stars that have measured effective temperatures consistent with $6250$\,K from the $1\sigma$ uncertainty. The dark black borders around the symbols are for systems with multiple transiting planets. Systems with measured true obliquities ($\psi$) are plotted as squares while projected obliquities ($\lambda$) are shown as circles. WASP-79b and HATS-3b from our previous publications \citep{2013ApJ...774L...9A,2014ApJ...792..112A} are also shown in this figure.
\\ [+1.5ex]
(A color version of this figure will be available in the online journal.)}
\label{Figure:obliquity_timescale}
\end{figure}

The most obvious outliers in Figure\,\ref{Figure:obliquity_temp} are the six planets labeled to left of the dashed red line at $T_\mathrm{eff}=6250$\,K, which are on significantly misaligned orbits. These systems clearly break the observed trend between obliquity and temperature. However, two of those planets (WASP-80b, \citealt{2013A&A...551A..80T}; and WASP-2b, \citealt{2011ApJ...738...50A}) do not have well constrained $\lambda$ values despite the small uncertainties reported. For WASP-80b, this is due to $\lambda$ being almost entirely dependent on the value of $v\sin i_{\star}$. For WASP-2b, \citet{2011ApJ...738...50A} could not detect the Rossiter--McLaughlin effect from the 66 spectra they obtained for this system that span the transit using the Planet Finding Spectrograph on Magellan and the High Dispersion Spectrograph on Subaru, contradicting the significant misalignment measured by \citet{2010A&A...524A..25T} from the 15 spectra they obtained using the High Accuracy Radial velocity Planet Searcher.

The sample of stars with measured spin--orbit angles still does appears to show that $\lambda$ has a weak dependence on host star temperature. Around $52\%\pm17\%$ of stars (out of 31 that have measured spin--orbit angles) with $T_\mathrm{eff}\geq6250$\,K host planets on misaligned orbits. This is in contrast to the $\sim\!32\%\pm12\%$ of stars with $T_\mathrm{eff}<6250$\,K (that have measured spin--orbit angles as previously mentioned) that host planets on misaligned orbits. Based on these statistics, high-obliquity orbits are more likely ($\gtrsim1\sigma$) to be found around stars $T_\mathrm{eff}\geq6250$\,K than stars $T_\mathrm{eff}<6250$\,K (a trend that has been noted by \citealt{2010ApJ...718L.145W}; \citealt{2012ApJ...757...18A}; and others).

The (weakly) observed temperature trend might be explained by the thickness of the stellar convective envelope and its ability to tidally dampen orbital obliquities \citep[as discussed in][]{2014ApJ...792..112A}. Stars with $T_\mathrm{eff}\geq6250$\,K have thin convective layers, therefore, planet--star tidal interactions are weak and unable to realign highly misaligned orbits \citep{2012ApJ...757...18A}. In contrast, stars with $T_\mathrm{eff}<6250$\,K have a thicker convective envelope, which results in stronger planet--star tidal interactions and shorter realignment timescales. It should be noted that while $T_\mathrm{eff}=6250$\,K is often used \citep[e.g.,][]{2010ApJ...718L.145W, 2012ApJ...757...18A} as the dividing boundary between stars with thin versus thick convective layers, this boundary is only approximate. The convective zone mass actually decreases exponentially as a function of stellar temperature above $T_\mathrm{eff}\sim5500$\,K while decreasing linearly below this temperature \citep{2001ApJ...556L..59P}.

Hence, the ability of a star to host planets on high-obliquity orbits for billions of years is dependent more on the tidal dissipation timescale than just the stellar effective temperature. Cool stars can host planets on misaligned orbits if the tidal dissipation timescale for realignment is very long \citep[i.e., as suggested by the realignment timescale Equations 2-4 in][]{2012ApJ...757...18A}. The relationship between tidal dissipation timescale and orbital obliquity is more clearly illustrated in Figure\,\ref{Figure:obliquity_timescale}, which is an updated plot\footnote{Figure\,\ref{Figure:obliquity_timescale} was produced from the compilation of stellar and planetary physical parameters as provided from \url{http://www.astro.keele.ac.uk/jkt/tepcat/allplanets-err.html}.} from \citet{2013ApJ...771...11A} and \citet{2014ApJ...792..112A} using Equations\,\ref{equat:convective_envelope} and \ref{equat:radiative_envelope}. It shows a trend toward higher orbital obliquities for longer relative tidal dissipation timescales. This trend does appear to be more obvious than the widely reported temperature versus obliquity correlation (see Figure\,\ref{Figure:obliquity_temp}). This is likely because high obliquities are dependent on other relevant physical parameters (i.e., $a/R_{\star}$, $M_{P}/M_{\star}$, and $M_{P}/M_{\star}$) than just stellar temperature alone.

\begin{equation}\label{equat:convective_envelope}
\frac{1}{\tau_{\mathrm{CE}}}=\frac{1}{10\times10^{9}\mathrm{yr}}q^{2}\left(\frac{a/R_{\star}}{40}\right)^{-6}
\end{equation}

\begin{equation}\label{equat:radiative_envelope}
\frac{1}{\tau_{\mathrm{RA}}}=\frac{1}{0.25\times5\times10^{9}\mathrm{yr}}q^{2}\left(1+q\right)^{5/6}\left(\frac{a/R_{\star}}{6}\right)^{-17/2}
\end{equation}

Where $\tau_{\mathrm{CE}}$ and $\tau_{\mathrm{RA}}$ are the tidal dissipation timescales considering either stars with convective ($T_\mathrm{eff}<6250$\,K) or radiative ($T_\mathrm{eff}\geq6250$\,K) envelopes, respectively, and $q$ is the planet-to-star mass ratio ($M_{P}/M_{\star}$). The tidal dissipation timescales $\tau_{CE}$ and $\tau_{RA}$ in Figure\,\ref{Figure:obliquity_timescale} were derived by \citet{2012ApJ...757...18A} from the studies of tidal friction in close binary stars, as carried out by \citet{1977A&A....57..383Z}. A potential caveat is the decision whether a star has convective or radiative envelope. \citet{2012ApJ...757...18A} has made the assumption that the convective envelope for stars $T_\mathrm{eff}\geq6250$\,K are too thin to realign orbital obliquities and therefore set the boundary at that temperature. However, this does not take into consideration the gradual thinning of the convective envelope with increasing stellar temperature.

To circumvent this issue, \citet{2012ApJ...757...18A} devised a second approach for calculating the tidal dissipation timescales by deriving an equation that takes into account the mass of the convective envelope ($\tau_{\mathrm{mcz}}$, see Equation\,\ref{equat:convective_mass}). 

\begin{equation}\label{equat:convective_mass}
\frac{1}{\tau_{\mathrm{mcz}}}=C \cdot M_{\mathrm{cz}} \cdot q^{2}\left(\frac{R}{a}\right)^{6}
\end{equation}

Where $M_{\mathrm{cz}}$ is the amount of mass in the convective envelope and $C$ is a proportionality constant equal to $10^{5}$\,g\,$\mathrm{s}^{-1}$ (as provided to us by private communication with Dr. Simon Albrecht, 2014). Dr. Simon Albrecht has also informed us that the equation for $\tau_{\mathrm{mcz}}$ (given as $1/\tau_{\mathrm{mcz}}\propto 1/M_{\mathrm{cz}}$ in the publication \citealt{2012ApJ...757...18A}) is incorrect and should be $1/\tau_{mcz}\propto M_{\mathrm{cz}}$. This correction has been implemented in Equation\,\ref{equat:convective_mass} and is the equation being used in this work. 

We present Table\,\ref{table:tidal_dissipation_timescales}, which list the $T_\mathrm{eff}$, age, $\lambda$, $\psi$ (if known), $M_{\mathrm{cz}}$, $\tau_{\mathrm{CE}}$ or $\tau_{\mathrm{RA}}$, and $\tau_{\mathrm{mcz}}$ and their associated uncertainties of the exoplanetary systems with measured spin--orbit angles\footnote{Spin--orbit alignments obtained from the Ren\'{e} Heller's Holt-Rossiter--McLaughlin Encyclopaedia on 2015 November 20. \url{http://www.astro.physik.uni-goettingen.de/~rheller/}.}, including WASP-66, WASP-87, and WASP-103 reported in this publication, as a reference for the community. We are not aware of any published summary of all these tidal dissipation timescales and the stellar convective zone masses. The stellar and planetary physical parameters used for calculating $M_{\mathrm{cz}}$, $\tau_{\mathrm{CE}}$ or $\tau_{\mathrm{RA}}$, and $\tau_{\mathrm{mcz}}$ were obtained from the TEPCat catalog\footnote{The TEPCat catalog can be accessed at \url{http://www.astro.keele.ac.uk/jkt/tepcat/tepcat.html}.} and the Extrasolar Planets Encyclopaedia.\footnote{The Extrasolar Planets Encyclopaedia can be accessed at \url{http://exoplanet.eu/}.} $M_{\mathrm{cz}}$ were calculated from the EZ-Web\footnote{The EZ-Web tool is available at the following web address: \url{http://www.astro.wisc.edu/~townsend/}.} tool made available by Richard Townsend.

Results from a direct imaging survey, carried out by \citet{2015ApJ...800..138N} to search for stellar companions around Hot Jupiters with measured spin--orbit alignments, shows no correlation between the incidence of stellar companions and planets being on misaligned or eccentric orbits. The Ngo et al. survey targeted a sample of 50 systems, 27 hosting planets on misaligned and/or eccentric orbits, and 23 on well aligned and circular orbits (classified as the ``control sample''), to determine if any correlations exist between misaligned/eccentric Hot Jupiters and the incidence of stellar companions. They discovered 19 stellar companions around 17 stars and from this result determined that the companion fraction for systems hosting misaligned planets is approximately equal (to within their reported uncertainties) to that of spin--orbit aligned systems. Their survey suggest that stellar companions may not play a dominant role in producing planets on high-obliquity orbits. Interestingly, \citet{2014ApJ...791...35L} found tentative evidence (at the $98$\% confidence) for stellar companions leading to an increased rate of close-in giant planets from a adaptive optics survey of 715 Kepler planetary system candidates with the Robo-AO robotic laser adaptive optics system. Therefore, migration mechanisms that require the presence of stellar companions to operate, such as the Kozai mechanism \citep{1962AJ.....67..591K,1962P&SS....9..719L,2007ApJ...669.1298F}, are disfavored. Other migration scenarios, such as planet-planet scattering \citep{2008ApJ...686..580C} or primordial disk misalignments \citep[from binary companions that were present at an earlier epoch but later removed through dynamical processes; see][]{2012Natur.491..418B}, might be more favored.

Our group is also conducting a similar direct imaging survey, but in the southern hemisphere, to search for stellar companions around planetary systems with measured spin--orbit alignments. We are using the Magellan Adaptive Optics (MagAO) system on the 6.5\,m Magellan Telescope at the Las Campanas Observatory in Chile to survey nearby stars within 250\,pc \citep[see][]{2014ApJ...792..112A,2014arXiv1403.0652A}. The results from our direct imaging search will be available soon and will complement the \citet{2015ApJ...800..138N} survey. Taken together, the results from the two surveys will provide a strong test of the Kozai mechanism as the dominant driver for misaligning planetary orbits.

In addition to these direct imaging searches, a radial velocity survey was conducted by \citet{2014ApJ...785..126K} of 51 transiting planetary systems (50 of which were in the \citealt{2015ApJ...800..138N} direct imaging survey). Knutson et al. found no correlation between planets on high-obliquity orbits and the occurrence rate of long-period massive planetary companions. This seems to suggest that planet--planet scattering may also not be playing a significant role in shaping the orbital obliquities of planets. \citet{2015ApJ...800..138N} offer an alternative hypothesis, proposing that the protoplanetary disks are perturbed out of alignment at the epoch of star and planet formation due to primordially bound stellar binaries. The binary companions are later removed from the planetary systems through dynamical interactions between stars in crowded stellar clusters, eliminating their current observational signatures \citep{2007MNRAS.378.1207M,2015ApJ...800..138N}.

If high-obliquity Hot Jupiters are indeed the result of primordial disk misalignments, then there should be a significant number of debris disks observed to be in misalignment with the spin axis of their host star. Recent observations of well aligned debris disk \citep[e.g.,][]{2013MNRAS.436..898K,2014MNRAS.438L..31G}, however, argue against this mechanism being responsible for the majority of spin--orbit misalignments. This model can also be tested from measurements of spin--orbit alignments in co-planar multi-planet systems with stellar companions. Since multiple transiting planet systems are nearly co-planar (at least for the planets that are transiting), presumably their orbits trace the plane of the protoplanetary disk from which the planets formed originally \citep{2013ApJ...771...11A}. Therefore, misalignments produced through dynamical mechanisms would result in planets on various orbital planes and with different observed obliquities. If these systems are found to be predominantly in spin--orbit alignment, then high-obliquity Hot Jupiter systems are likely the result of high eccentricity migration (i.e., planet-planet scatterings, secular chaos, Kozai-Lidov resonances, etc.) instead of primordial star--disk misalignments \citep{2012ApJ...757...18A}. If, however, the obliquities of multi-planet systems are similarly distributed to those observed for Hot Jupiters, then this would suggest that spin--orbit misalignments are produced through processes other than migration such as primordial disk misalignments. Recent spin--orbit alignment measurements of transiting multi-planet systems have revealed that six host planets on low-obliquity orbits \citep{2012Natur.487..449S,2013ApJ...771...11A,2013ApJ...766..101C,2015ApJ...812L..11S}, while one host planets on misaligned orbits \citep{2013Sci...342..331H}. Statistical studies on the distribution of orbital obliquities of Kepler transiting planets, using the photometric stellar rotational amplitudes produced from star spots in the Kepler light curves, suggest that planets are more likely to be found on misaligned orbits around hot stars than cool stars \citep{2015ApJ...801....3M} and if they orbit a star with only a single detected transiting planet as opposed to orbiting a star with multiple detected transiting planets \citep{2014ApJ...796...47M}. The sample size, however, is too small to draw any definitive conclusions. Therefore, more multi-planet systems need to have their spin--orbit angles measured to test the primordial disk misalignments hypothesis and to resolve the migration processes of high-obliquity planets. Such a sample of suitably bright stars with multiple transiting planets should be found by the ongoing \textit{K}2 mission, the new two-wheel operation mode of \textit{Kepler}, in the near future \citep[e.g., see][]{2014PASP..126..398H}.

\section{Conclusions}\label{sec:Conclusions_wasp-66-103}
Measurements of the spin--orbit angles of WASP-66b, WASP-87b, and WASP-103b from the work presented in this paper, taken together with the whole sample of systems, seems to support the \citet{2010ApJ...718L.145W} and \citet{2012ApJ...757...18A} hypothesis that Hot Jupiters once had a broad distribution of orbital obliquities (from the migration mechanisms that produced them). In addition, the data suggest that high eccentricity migration models are slightly more favored over disk migration models. This is because planets migrate too quickly (according to the current disk migrations models) and misaligned debris disk appear to be rare (though the sample size is still too small to draw definitive conclusions). Alternatively, high obliquities could be produced from a combination of two or more mechanisms operating concurrently \citep[e.g., see][]{2011ApJ...729..138M}. It is even possible that there is no `one' dominant mechanism common to all Hot Jupiter systems, implying different mechanisms are operating in each system and this is dependent on the initial formation conditions of the system. Therefore, we believe a lingering question still remains: what are the dominant mechanism(s) producing high obliquity Hot Jupiters?

Resolving this question will require a further expansion of the sample of systems with measured spin--orbit angles, in particular: low-mass planets, long-period planets, and multi-planet systems. Finally, direct imaging \citep[e.g.,][]{2014arXiv1403.0652A,2015ApJ...800..138N} and radial velocity \citep[e.g.,][]{2014ApJ...785..126K} searches for stellar and planetary companions around stars hosting Hot Jupiters with spin--orbit angle measurements will empirically test Kozai resonances (one of the most commonly proposed migration mechanisms) as the dominant driving force behind the production of the large number of Hot Jupiters on high-obliquity orbits.

\makeatletter
\def\blfootnote{\gdef\@thefnmark{}\@footnotetext}
\makeatother

\blfootnote{\textit{Note added in proof.} Repeating the analysis of Section\,\ref{sec:RM_analysis} using updated parameters for WASP-103 published while this paper was being refereed \citep{2015MNRAS.447..711S} results in a spin-orbit angle and $v\sin i_{\star}$ consistent with those in Table\,\ref{table:WASP-103_Parameters} ($5^{\circ} \pm 38^{\circ}$ and $8.3 \pm 0.8$\,\kms).}

\acknowledgments

The research work presented in the paper at UNSW has been supported by ARC Australian Research Council grants DP0774000, DP130102695, LE0989347, and FS100100046. Work at the Australian National University is supported by ARC Laureate Fellowship Grant FL0992131. We thank Vincent Dumont for his helpful comments and suggestions on this manuscript. We also thank Dr. Yanan Fan for her helpful comments on implementing a Markov Chain Monte Carlo model in ExOSAM. We acknowledge the use of the SIMBAD database, operated at CDS, Strasbourg, France. This research has made use of NASA's Astrophysics Data System, the Ren\'{e} Heller's Holt-Rossiter--McLaughlin Encyclopaedia (\url{http://www.astro.physik.uni-goettingen.de/~rheller/}), the Exoplanet Orbit Database and the Exoplanet Data Explorer at \url{exoplanets.org}, and the Extrasolar Planets Encyclopaedia at \url{http://exoplanet.eu}.

\bibliography{WASP-66_WASP-87_WASP-103}

\clearpage
\begin{table*}
  \centering
  \caption{The tidal-dissipation timescales of the sample of transiting planetary systems with measured spin--orbit angles.}
  \label{table:tidal_dissipation_timescales}
  \resizebox{\textwidth}{!}{%
    \begin{tabular}{l c c c c c c c c c c}
      \hline\hline \\ [-2.0ex]
      \multirow{2}{*}{System} &
      $T_{eff}$ &
      Age &
      $\lambda$ &
      $\psi$ &
      $M_{cz}$ &
      $\tau_{\mathrm{RA\, or\, CE}}$ &
      \multirow{2}{*}{$\frac{\tau_{\mathrm{RA\, or\, CE}}(\mathrm{year})}{5\times10^{9}(\mathrm{year})}$} & 
      $\tau_{mcz}$ &
      \multirow{2}{*}{$\frac{\tau_{mcz}(\mathrm{year})}{\mathrm{Age}(\mathrm{year})}$} &
      \multirow{2}{*}{References} \\
      & (K) & (Gyr) & (deg) & (deg) & ($M_{\odot}$) & (year) &  & (year) &  & \\ [0.5ex]
      \hline \\ [-2.0ex]
      55 Cnc e & $5196\pm24$ & $10.2\pm2.5$ & $72.4^{+23.7}_{-11.5}$ & \nodata & $5.09\times10^{-2}$ & $(6.51 \pm 0.42) \times 10^{12}$ & $(1.30 \pm 0.08) \times 10^{3}$ & $(5.24 \pm 0.43) \times 10^{8}$ & $(5.14 \pm 1.33) \times 10^{-2}$ & 1,2,3 \\ [1.5ex]
      CoRoT-01 & $5950\pm150$ & $^{\mathrm{a}}0.5$ & $77\pm11$ & \nodata & $2.11\times10^{-2}$ & $(2.86 \pm 0.58) \times 10^{10}$ & $(5.71 \pm 1.16) \times 10^{0}$ & $(5.55 \pm 1.32) \times 10^{6}$ & $(1.11 \pm 0.26) \times 10^{-2}$ & 4,5 \\ [1.5ex]
      CoRoT-02 & $5598\pm50$ & $^{\mathrm{a}}0.5$ & $7.2\pm4.5$ & \nodata & $3.45\times10^{-3}$ & $(2.00 \pm 0.44) \times 10^{10}$ & $(4.00 \pm 0.89) \times 10^{0}$ & $(2.37 \pm 0.55) \times 10^{7}$ & $(4.75 \pm 1.09) \times 10^{-2}$ & 6,7 \\ [1.5ex]
      CoRoT-03 & $6740\pm140$ & $2.2\pm0.6$ & $37.6^{+10}_{-22.3}$ & \nodata & $1.16\times10^{-4}$ & $(5.68 \pm 5.47) \times 10^{13}$ & $(1.14 \pm 1.09) \times 10^{4}$ & $(9.33 \pm 6.77) \times 10^{7}$ & $(4.24 \pm 3.29) \times 10^{-2}$ & 8,5 \\ [1.5ex]
      CoRoT-11 & $6440\pm120$ & $2\pm1$ & $0.1\pm2.6$ & \nodata & $8.99\times10^{-6}$ & $(1.28 \pm 0.86) \times 10^{15}$ & $(2.55 \pm 1.72) \times 10^{5}$ & $(3.77 \pm 1.95) \times 10^{10}$ & $(1.88 \pm 1.35) \times 10^{1}$ & 9,5 \\ [1.5ex]
      CoRoT-18 & $5440\pm100$ & $^{\mathrm{a}}0.5$ & $-10\pm20$ & $20\pm20$ & $3.07\times10^{-2}$ & $(1.61 \pm 0.47) \times 10^{10}$ & $(3.22 \pm 0.93) \times 10^{0}$ & $(2.15 \pm 0.71) \times 10^{6}$ & $(4.30 \pm 1.42) \times 10^{-3}$ & 10,7 \\ [1.5ex]
      CoRoT-19 & $6090\pm70$ & $5\pm1$ & $-52^{+27}_{-22}$ & \nodata & $1.18\times10^{-2}$ & $(3.65 \pm 0.59) \times 10^{11}$ & $(7.29 \pm 1.18) \times 10^{1}$ & $(1.27 \pm 0.49) \times 10^{8}$ & $(2.53 \pm 1.10) \times 10^{-2}$ & 11,7 \\ [1.5ex]
      HAT-P-01 & $5975\pm50$ & $3.6$ & $3.7\pm2.1$ & \nodata & $5.05\times10^{-3}$ & $(1.43 \pm 0.08) \times 10^{13}$ & $(2.87 \pm 0.16) \times 10^{3}$ & $(1.16 \pm 0.10) \times 10^{10}$ & $(3.23 \pm 0.28) \times 10^{0}$ & 12,13 \\ [1.5ex]
      HAT-P-02 & $6290\pm60$ & $2.7\pm0.5$ & $0.2^{+12.2}_{-12.5}$ & \nodata & $1.03\times10^{-3}$ & $(6.38 \pm 2.62) \times 10^{14}$ & $(1.28 \pm 0.52) \times 10^{5}$ & $(9.39 \pm 3.51) \times 10^{7}$ & $(3.48 \pm 1.45) \times 10^{-2}$ & 14,15 \\ [1.5ex]
      HAT-P-04 & $5860\pm80$ & $4.2\pm0.6$ & $-4.9\pm11.9$ & \nodata & $1.19\times10^{-2}$ & $(4.36 \pm 0.40) \times 10^{11}$ & $(8.72 \pm 0.80) \times 10^{1}$ & $(1.50 \pm 0.24) \times 10^{8}$ & $(3.57 \pm 0.77) \times 10^{-2}$ & 16,5 \\ [1.5ex]
      HAT-P-06 & $6570\pm80$ & $2.3\pm0.7$ & $165\pm6$ & \nodata & $3.70\times10^{-6}$ & $(1.25 \pm 0.17) \times 10^{16}$ & $(2.49 \pm 0.34) \times 10^{6}$ & $(7.38 \pm 1.16) \times 10^{11}$ & $(3.21 \pm 1.10) \times 10^{2}$ & 17,18 \\ [1.5ex]
      HAT-P-07 & $6350\pm80$ & $2.07^{+0.28}_{-0.23}$ & $155\pm37$ & $101\pm2$ or $87\pm2$ & $7.08\times10^{-4}$ & $(4.51 \pm 0.30) \times 10^{13}$ & $(9.02 \pm 0.60) \times 10^{3}$ & $(5.86 \pm 0.35) \times 10^{7}$ & $(2.83 \pm 0.39) \times 10^{-2}$ & 17,5,119 \\ [1.5ex]
      HAT-P-08 & $6200\pm80$ & $3.4\pm1$ & $-17^{+9.2}_{-11.5}$ & \nodata & $3.14\times10^{-3}$ & $(1.61 \pm 0.14) \times 10^{11}$ & $(3.21 \pm 0.28) \times 10^{1}$ & $(2.09 \pm 0.24) \times 10^{8}$ & $(6.16 \pm 1.94) \times 10^{-2}$ & 19,20 \\ [1.5ex]
      HAT-P-09 & $6350\pm150$ & $1.6\pm1.4$ & $-16\pm8$ & \nodata & $3.02\times10^{-4}$ & $(7.11 \pm 1.71) \times 10^{16}$ & $(1.42 \pm 0.34) \times 10^{7}$ & $(3.69 \pm 0.88) \times 10^{10}$ & $(2.31 \pm 2.09) \times 10^{1}$ & 19,7 \\ [1.5ex]
      HAT-P-11 & $4780\pm50$ & $6.5\pm4.1$ & $103^{+26}_{-10}$ & \nodata & $6.45\times10^{-2}$ & $(4.64 \pm 0.54) \times 10^{15}$ & $(9.27 \pm 1.07) \times 10^{5}$ & $(2.94 \pm 0.38) \times 10^{11}$ & $(4.53 \pm 2.92) \times 10^{1}$ & 21,5 \\ [1.5ex]
      HAT-P-13 & $5653\pm90$ & $5\pm0.8$ & $1.9\pm8.6$ & \nodata & $^{\mathrm{b}}6.74\times10^{-3}$ & $(1.36 \pm 0.08) \times 10^{11}$ & $(2.71 \pm 0.16) \times 10^{1}$ & $^{\mathrm{b}}(8.24 \pm 0.77) \times 10^{7}$ & $(1.65 \pm 0.31) \times 10^{-2}$ & 22,23 \\ [1.5ex]
      HAT-P-14 & $6600\pm90$ & $1.3\pm0.4$ & $189.1\pm5.1$ & \nodata & $4.27\times10^{-6}$ & $(8.02 \pm 1.07) \times 10^{15}$ & $(1.60 \pm 0.21) \times 10^{6}$ & $(3.16 \pm 0.42) \times 10^{11}$ & $(2.43 \pm 0.82) \times 10^{2}$ & 16,7 \\ [1.5ex]
      HAT-P-16 & $6140\pm72$ & $2\pm0.8$ & $-10\pm16$ & \nodata & $1.08\times10^{-3}$ & $(4.57 \pm 0.71) \times 10^{10}$ & $(9.15 \pm 1.42) \times 10^{0}$ & $(1.73 \pm 0.29) \times 10^{8}$ & $(8.67 \pm 3.76) \times 10^{-2}$ & 19,24 \\ [1.5ex]
      HAT-P-17 & $5246\pm80$ & $7.8\pm3.3$ & $19^{+14}_{-16}$ & \nodata & $4.59\times10^{-2}$ & $(9.25 \pm 0.53) \times 10^{14}$ & $(1.85 \pm 0.11) \times 10^{5}$ & $(8.26 \pm 0.76) \times 10^{10}$ & $(1.06 \pm 0.46) \times 10^{1}$ & 25,26 \\ [1.5ex]
      HAT-P-18 & $4870\pm50$ & $12.4^{+4.4}_{-6.4}$ & $132\pm15$ & \nodata & $5.76\times10^{-2}$ & $(9.16 \pm 0.56) \times 10^{14}$ & $(1.83 \pm 0.11) \times 10^{5}$ & $(6.51 \pm 0.73) \times 10^{10}$ & $(5.25 \pm 2.36) \times 10^{0}$ & 27 \\ [1.5ex]
      HAT-P-23 & $5905\pm80$ & $4\pm1$ & $15\pm22$ & \nodata & $6.02\times10^{-3}$ & $(3.97 \pm 0.64) \times 10^{9}$ & $(7.95 \pm 1.27) \times 10^{-1}$ & $(2.70 \pm 0.59) \times 10^{6}$ & $(6.76 \pm 2.25) \times 10^{-4}$ & 19,28 \\ [1.5ex]
      HAT-P-24 & $6373\pm80$ & $2.8\pm0.6$ & $20\pm16$ & \nodata & $1.37\times10^{-4}$ & $(3.50 \pm 0.36) \times 10^{16}$ & $(7.00 \pm 0.72) \times 10^{6}$ & $(5.10 \pm 0.76) \times 10^{10}$ & $(1.82 \pm 0.47) \times 10^{1}$ & 17,29 \\ [1.5ex]
      HAT-P-27 & $5316\pm55$ & $4.4^{+3.8}_{-2.6}$ & $24.2^{+76}_{-44.5}$ & \nodata & $4.38\times10^{-2}$ & $(4.43 \pm 0.39) \times 10^{12}$ & $(8.85 \pm 0.79) \times 10^{2}$ & $(4.14 \pm 0.65) \times 10^{8}$ & $(9.41 \pm 7.00) \times 10^{-2}$ & 30,31 \\ [1.5ex]
      HAT-P-30 & $6338\pm42$ & $1^{+0.8}_{-0.5}$ & $73.5\pm9$ & \nodata & $1.13\times10^{-3}$ & $(2.52 \pm 0.23) \times 10^{16}$ & $(5.05 \pm 0.45) \times 10^{6}$ & $(4.92 \pm 0.61) \times 10^{9}$ & $(4.92 \pm 3.25) \times 10^{0}$ & 32 \\ [1.5ex]
      HAT-P-32 & $6207\pm88$ & $2.7\pm0.8$ & $85\pm1.5$ & \nodata & $2.52\times10^{-4}$ & $(2.39 \pm 0.80) \times 10^{11}$ & $(4.77 \pm 1.60) \times 10^{1}$ & $(3.88 \pm 1.31) \times 10^{9}$ & $(1.44 \pm 0.65) \times 10^{0}$ & 17,33 \\ [1.5ex]
      HAT-P-34 & $6442\pm88$ & $1.7\pm0.5$ & $0\pm14$ & \nodata & $2.09\times10^{-4}$ & $(1.17 \pm 0.37) \times 10^{16}$ & $(2.34 \pm 0.74) \times 10^{6}$ & $(6.67 \pm 2.01) \times 10^{9}$ & $(3.92 \pm 1.65) \times 10^{0}$ & 17,34 \\ [1.5ex]
      HAT-P-36 & $5620\pm40$ & $4.5^{+3.1}_{-3.9}$ & $-14\pm18$ & $25^{+38}_{-25}$ & $2.35\times10^{-2}$ & $(1.19 \pm 0.18) \times 10^{10}$ & $(2.38 \pm 0.36) \times 10^{0}$ & $(2.08 \pm 0.33) \times 10^{6}$ & $(4.61 \pm 3.66) \times 10^{-4}$ & 35,34 \\ [1.5ex]
      HATS-2 & $5227\pm95$ & $9.7\pm2.9$ & $8\pm8$ & \nodata & $4.52\times10^{-2}$ & $(3.21 \pm 0.92) \times 10^{10}$ & $(6.42 \pm 1.85) \times 10^{0}$ & $(2.91 \pm 0.86) \times 10^{6}$ & $(3.00 \pm 1.26) \times 10^{-4}$ & 36 \\ [1.5ex]
      HATS-3 & $6351\pm76$ & $3.2$ & $3\pm25$ & \nodata & $4.19\times10^{-5}$ & $(1.07 \pm 0.35) \times 10^{16}$ & $(2.14 \pm 0.69) \times 10^{6}$ & $(5.59 \pm 1.38) \times 10^{10}$ & $(1.75 \pm 0.43) \times 10^{1}$ & 37,38 \\ [1.5ex]
      HATS-14 & $5408\pm65$ & $4.9\pm1.7$ & $76^{+4}_{-5}$ & \nodata & $3.28\times10^{-2}$ & $(1.03 \pm 0.04) \times 10^{12}$ & $(2.07 \pm 0.08) \times 10^{2}$ & $(1.29 \pm 0.09) \times 10^{8}$ & $(2.64 \pm 0.93) \times 10^{-2}$ & 39 \\ [1.5ex]
      HD 017156 & $6079\pm56$ & $3.37^{+0.24}_{-0.2}$ & $10\pm5.1$ & \nodata & $4.85\times10^{-3}$ & $(7.44 \pm 0.99) \times 10^{13}$ & $(1.49 \pm 0.20) \times 10^{4}$ & $(6.28 \pm 0.98) \times 10^{10}$ & $(1.86 \pm 0.32) \times 10^{1}$ & 40,5 \\ [1.5ex]
      HD 080606 & $5574\pm72$ & $1.6^{+1.8}_{-1.1}$ & $42\pm8$ & \nodata & $3.72\times10^{-2}$ & $(1.18 \pm 0.26) \times 10^{17}$ & $(2.35 \pm 0.52) \times 10^{7}$ & $(1.30 \pm 0.31) \times 10^{13}$ & $(8.09 \pm 7.59) \times 10^{3}$ & 41,5,42 \\ [1.5ex]
      HD 149026 & $6147\pm50$ & $2\pm0.8$ & $12\pm7$ & \nodata & $7.39\times10^{-4}$ & $(1.70 \pm 0.09) \times 10^{12}$ & $(3.39 \pm 0.19) \times 10^{2}$ & $(9.41 \pm 0.89) \times 10^{9}$ & $(4.70 \pm 1.93) \times 10^{0}$ & 17,43 \\ [1.5ex]
      HD 189733 & $5050\pm50$ & $0.6$ & $-0.5\pm0.3$ & \nodata & $3.71\times10^{-2}$ & $(7.50 \pm 0.62) \times 10^{11}$ & $(1.50 \pm 0.12) \times 10^{2}$ & $(8.28 \pm 1.01) \times 10^{7}$ & $(1.38 \pm 0.17) \times 10^{-1}$ & 44,15 \\ [1.5ex]
      HD 209458 & $6117\pm50$ & $4.5$ & $-4.4\pm1.4$ & \nodata & $7.48\times10^{-3}$ & $(3.18 \pm 0.13) \times 10^{12}$ & $(6.35 \pm 0.26) \times 10^{2}$ & $(1.74 \pm 0.10) \times 10^{9}$ & $(3.86 \pm 0.22) \times 10^{-1}$ & 45,15 \\ [1.5ex]
      KELT-1 & $6516\pm49$ & $1.75\pm0.25$ & $2\pm16$ & \nodata & $1.13\times10^{-4}$ & $(4.30 \pm 5.74) \times 10^{10}$ & $(8.59 \pm 11.50) \times 10^{0}$ & $(5.13 \pm 5.07) \times 10^{5}$ & $(2.93 \pm 2.93) \times 10^{-4}$ & 46 \\ [1.5ex]
      Kepler-08 & $6213\pm150$ & $3.84\pm1.5$ & $5\pm7$ & \nodata & $1.29\times10^{-3}$ & $(1.34 \pm 0.43) \times 10^{12}$ & $(2.68 \pm 0.86) \times 10^{2}$ & $(4.25 \pm 1.40) \times 10^{9}$ & $(1.11 \pm 0.57) \times 10^{0}$ & 17,5 \\ [1.5ex]
      Kepler-13 & $7650\pm250$ & $0.708^{+0.183}_{-0.146}$ & $58.6\pm2$ & $60\pm2$ & $1.00\times10^{-8}$ & $(1.38 \pm 4.79) \times 10^{14}$ & $(2.76 \pm 9.58) \times 10^{4}$ & $(4.86 \pm 12.50) \times 10^{12}$ & $(6.86 \pm 17.70) \times 10^{3}$ & 47,48,49,119 \\ [1.5ex]
      Kepler-17 & $5781\pm85$ & $1.78$ & $0\pm15$ & $0\pm15$ & $2.52\times10^{-2}$ & $(1.90 \pm 0.46) \times 10^{10}$ & $(3.80 \pm 0.93) \times 10^{0}$ & $(3.09 \pm 0.83) \times 10^{6}$ & $(1.73 \pm 0.47) \times 10^{-3}$ & 50,7 \\ [1.5ex]
      Kepler-25c & $6270\pm79$ & $^{\mathrm{a}}0.5$ & $-0.5\pm5.7$ & \nodata & $2.40\times10^{-4}$ & $(3.81 \pm 1.71) \times 10^{21}$ & $(7.62 \pm 3.42) \times 10^{11}$ & $(3.78 \pm 1.26) \times 10^{14}$ & $(7.55 \pm 2.52) \times 10^{5}$ & 51,52 \\ [1.5ex]
      Kepler-30b & $5498\pm54$ & $2\pm0.1$ & $4\pm10$ & \nodata & $2.92\times10^{-2}$ & $(9.51 \pm 1.74) \times 10^{18}$ & $(1.90 \pm 0.35) \times 10^{9}$ & $(1.33 \pm 0.47) \times 10^{15}$ & $(6.67 \pm 2.40) \times 10^{5}$ & 53 \\ [1.5ex]
      Kepler-30c & $5498\pm54$ & $2\pm0.1$ & $4\pm10$ & \nodata & $2.92\times10^{-2}$ & $(6.36 \pm 1.65) \times 10^{16}$ & $(1.27 \pm 0.33) \times 10^{7}$ & $(8.92 \pm 3.57) \times 10^{12}$ & $(4.46 \pm 1.80) \times 10^{3}$ & 54 \\ [1.5ex]
      Kepler-30d & $5498\pm54$ & $2\pm0.1$ & $4\pm10$ & \nodata & $2.92\times10^{-2}$ & $(1.04 \pm 0.18) \times 10^{21}$ & $(2.08 \pm 0.36) \times 10^{11}$ & $(1.46 \pm 0.51) \times 10^{17}$ & $(7.31 \pm 2.59) \times 10^{7}$ & 55 \\ [1.5ex]
      Kepler-50b & $6225\pm66$ & $3.8\pm0.8$ & $10\pm0$ & $10\pm0$ & $7.00\times10^{-3}$ & $(5.44 \pm 3.89) \times 10^{14}$ & $(1.09 \pm 0.78) \times 10^{5}$ & $(3.19 \pm 2.78) \times 10^{11}$ & $(8.38 \pm 7.53) \times 10^{1}$ & 54 \\ [1.5ex]
      Kepler-50c & $6225\pm66$ & $3.8\pm0.8$ & $10\pm0$ & $10\pm0$ & $7.00\times10^{-3}$ & $(9.36 \pm 6.69) \times 10^{14}$ & $(1.87 \pm 1.34) \times 10^{5}$ & $(5.48 \pm 4.83) \times 10^{11}$ & $(1.44 \pm 1.31) \times 10^{2}$ & 54 \\ [1.5ex]
      Kepler-56b & $4840\pm97$ & $3.5\pm1.3$ & $47\pm6$ & $47\pm6$ & $6.73\times10^{-3}$ & $(1.99 \pm 0.47) \times 10^{13}$ & $(3.98 \pm 0.94) \times 10^{3}$ & $(1.21 \pm 0.32) \times 10^{10}$ & $(3.47 \pm 1.58) \times 10^{0}$ & 55 \\ [1.5ex]
      Kepler-56c & $4840\pm97$ & $3.5\pm1.3$ & $47\pm6$ & $47\pm6$ & $6.73\times10^{-3}$ & $(5.05 \pm 0.87) \times 10^{12}$ & $(1.01 \pm 0.17) \times 10^{3}$ & $(3.07 \pm 0.65) \times 10^{9}$ & $(8.78 \pm 3.75) \times 10^{-1}$ & 55 \\ [1.5ex]
      Kepler-63 & $5576\pm50$ & $0.21\pm0.045$ & $-110^{+22}_{-14}$ & $145^{+9}_{-14}$ & $3.73\times10^{-2}$ & $(8.68 \pm 6.58) \times 10^{14}$ & $(1.74 \pm 1.32) \times 10^{5}$ & $(9.54 \pm 7.28) \times 10^{10}$ & $(4.54 \pm 3.60) \times 10^{2}$ & 56 \\ [1.5ex]
      Kepler-65b & $6211\pm66$ & $2.9\pm0.7$ & $10\pm0$ & $10\pm0$ & $9.94\times10^{-4}$ & $(9.66 \pm 6.83) \times 10^{14}$ & $(1.93 \pm 1.37) \times 10^{5}$ & $(3.98 \pm 2.85) \times 10^{12}$ & $(1.37 \pm 1.04) \times 10^{3}$ & 54 \\ [1.5ex]
      Kepler-65c & $6211\pm66$ & $2.9\pm0.7$ & $10\pm0$ & $10\pm0$ & $9.94\times10^{-4}$ & $(7.40 \pm 7.67) \times 10^{14}$ & $(1.48 \pm 1.53) \times 10^{5}$ & $(3.05 \pm 3.18) \times 10^{12}$ & $(1.05 \pm 1.13) \times 10^{3}$ & 54 \\ [1.5ex]
      Kepler-65d & $6211\pm66$ & $2.9\pm0.7$ & $10\pm0$ & $10\pm0$ & $9.94\times10^{-4}$ & $(2.05 \pm 1.45) \times 10^{16}$ & $(4.10 \pm 2.90) \times 10^{6}$ & $(8.45 \pm 6.08) \times 10^{13}$ & $(2.91 \pm 2.21) \times 10^{4}$ & 54 \\ [1.5ex]
      Kepler-89b & $6182\pm82$ & $3.9^{+0.3}_{-0.2}$ & $-6^{+13}_{-11}$ & \nodata & $5.21\times10^{-3}$ & $(5.77 \pm 3.47) \times 10^{14}$ & $(1.15 \pm 0.69) \times 10^{5}$ & $(4.54 \pm 2.91) \times 10^{11}$ & $(1.16 \pm 0.75) \times 10^{2}$ & 57,58 \\ [1.5ex]
      Kepler-89d & $6182\pm82$ & $3.9^{+0.3}_{-0.2}$ & $-7^{+13}_{-11}$ & \nodata & $5.21\times10^{-3}$ & $(7.19 \pm 1.14) \times 10^{15}$ & $(1.44 \pm 0.23) \times 10^{6}$ & $(5.65 \pm 1.55) \times 10^{12}$ & $(1.45 \pm 0.41) \times 10^{3}$ & 57,58 \\ [1.5ex]
      KOI-12 & $6820\pm120$ & $1.5\pm0.5$ & $12.5^{+3}_{-2.9}$ & \nodata & $2.40\times10^{-7}$ & $(1.02 \pm 0.05) \times 10^{18}$ & $(2.04 \pm 0.09) \times 10^{8}$ & $(7.94 \pm 1.83) \times 10^{13}$ & $(5.29 \pm 2.15) \times 10^{4}$ & 59 \\ [1.5ex]
      KOI-1257 & $5520\pm80$ & $9.3\pm3$ & $74^{+32}_{-46}$ & \nodata & $3.55\times10^{-2}$ & $(1.84 \pm 1.12) \times 10^{17}$ & $(3.68 \pm 2.23) \times 10^{7}$ & $(2.12 \pm 1.44) \times 10^{13}$ & $(2.28 \pm 1.71) \times 10^{3}$ & 60 \\ [1.5ex]
      Qatar-1 & $4910\pm100$ & $4$ & $-8.4\pm7.1$ & \nodata & $6.42\times10^{-2}$ & $(6.81 \pm 0.69) \times 10^{10}$ & $(1.36 \pm 0.14) \times 10^{1}$ & $(4.35 \pm 0.79) \times 10^{6}$ & $(1.09 \pm 0.20) \times 10^{-3}$ & 61 \\ [1.5ex]
      \hline \\ [-2.0ex]
    \end{tabular}}
\end{table*}

\hfill
\begin{table*}
  \centering
  \resizebox{\textwidth}{!}{%
      \begin{tabular}{l c c c c c c c c c c}
        \multicolumn{11}{c}{\large \textbf{Table 6} -- \emph{Continued}}\\ [2ex]
        \hline\hline \\ [-2.0ex]
        \multirow{2}{*}{System} &
        $T_{eff}$ &
        Age &
        $\lambda$ &
        $\psi$ &
        $M_{cz}$ &
        $\tau_{\mathrm{RA\, or\, CE}}$ &
        \multirow{2}{*}{$\frac{\tau_{\mathrm{RA\, or\, CE}}(\mathrm{year})}{5\times10^{9}(\mathrm{year})}$} & 
        $\tau_{mcz}$ &
        \multirow{2}{*}{$\frac{\tau_{mcz}(\mathrm{year})}{\mathrm{Age}(\mathrm{year})}$} &
        \multirow{2}{*}{References} \\
        & (K) & (Gyr) & (deg) & (deg) & ($M_{\odot}$) & (year) &  & (year) &  & \\ [0.5ex]
        \hline \\ [-2.0ex]
        TrES-1 & $5226\pm50$ & $2.5\pm1.4$ & $30\pm21$ & \nodata & $4.71\times10^{-2}$ & $(4.57 \pm 0.57) \times 10^{12}$ & $(9.15 \pm 1.15) \times 10^{2}$ & $(3.98 \pm 0.58) \times 10^{8}$ & $(1.59 \pm 0.92) \times 10^{-1}$ & 62,15 \\ [1.5ex]
        TrES-2 & $5850\pm50$ & $5.1^{+2.7}_{-2.3}$ & $-9\pm12$ & \nodata & $6.96\times10^{-3}$ & $(4.58 \pm 0.42) \times 10^{11}$ & $(9.16 \pm 0.84) \times 10^{1}$ & $(2.70 \pm 0.30) \times 10^{8}$ & $(5.29 \pm 2.66) \times 10^{-2}$ & 63,5 \\ [1.5ex]
        TrES-4 & $6200\pm75$ & $2.9\pm0.3$ & $6.3\pm4.7$ & \nodata & $2.81\times10^{-3}$ & $(2.48 \pm 0.36) \times 10^{11}$ & $(4.95 \pm 0.71) \times 10^{1}$ & $(3.61 \pm 0.69) \times 10^{8}$ & $(1.24 \pm 0.27) \times 10^{-1}$ & 64,7 \\ [1.5ex]
        WASP-01 & $6160\pm64$ & $2\pm1$ & $-59^{+99}_{-26}$ & \nodata & $1.04\times10^{-3}$ & $(1.91 \pm 0.21) \times 10^{11}$ & $(3.82 \pm 0.42) \times 10^{1}$ & $(7.52 \pm 0.94) \times 10^{8}$ & $(3.76 \pm 1.94) \times 10^{-1}$ & 65,66,67 \\ [1.5ex]
        WASP-02 & $5170\pm60$ & $^{\mathrm{a}}0.5$ & $-153^{+15}_{-11}$ & \nodata & $5.30\times10^{-2}$ & $(6.95 \pm 0.62) \times 10^{11}$ & $(1.39 \pm 0.12) \times 10^{2}$ & $(5.37 \pm 0.61) \times 10^{7}$ & $(1.07 \pm 0.12) \times 10^{-1}$ & 68,7 \\ [1.5ex]
        WASP-03 & $6340\pm90$ & $2.1\pm1.4$ & $5^{+6}_{-5}$ & \nodata & $7.58\times10^{-3}$ & $(1.25 \pm 0.25) \times 10^{14}$ & $(2.49 \pm 0.51) \times 10^{4}$ & $(9.45 \pm 1.75) \times 10^{6}$ & $(4.50 \pm 3.11) \times 10^{-3}$ & 69,70,71 \\ [1.5ex]
        WASP-04 & $5540\pm55$ & $5.2^{+3.8}_{-3.2}$ & $4^{+34}_{-43}$ & \nodata & $1.45\times10^{-2}$ & $(3.98 \pm 0.40) \times 10^{10}$ & $(7.95 \pm 0.79) \times 10^{0}$ & $(1.12 \pm 0.14) \times 10^{7}$ & $(2.16 \pm 1.48) \times 10^{-3}$ & 68,7,72 \\ [1.5ex]
        WASP-05 & $5770\pm65$ & $3\pm1.4$ & $12.1^{+8}_{-10}$ & \nodata & $2.33\times10^{-2}$ & $(2.83 \pm 0.24) \times 10^{10}$ & $(5.66 \pm 0.49) \times 10^{0}$ & $(4.97 \pm 0.64) \times 10^{6}$ & $(1.66 \pm 0.80) \times 10^{-3}$ & 68,7 \\ [1.5ex]
        WASP-06 & $5375\pm65$ & $11\pm7$ & $-11^{+18}_{-14}$ & \nodata & $1.75\times10^{-2}$ & $(1.04 \pm 0.10) \times 10^{13}$ & $(2.08 \pm 0.20) \times 10^{3}$ & $(2.43 \pm 0.34) \times 10^{9}$ & $(2.21 \pm 1.44) \times 10^{-1}$ & 73 \\ [1.5ex]
        WASP-07 & $6520\pm70$ & $2.4^{+0.8}_{-1}$ & $86\pm6$ & \nodata & $2.67\times10^{-4}$ & $(8.34 \pm 2.65) \times 10^{16}$ & $(1.67 \pm 0.53) \times 10^{7}$ & $(4.15 \pm 1.16) \times 10^{10}$ & $(1.73 \pm 0.81) \times 10^{1}$ & 74,7 \\ [1.5ex]
        WASP-08 & $5600\pm80$ & $4\pm1$ & $-123.3^{+3.4}_{-4.4}$ & \nodata & $1.60\times10^{-2}$ & $(2.05 \pm 0.25) \times 10^{13}$ & $(4.11 \pm 0.50) \times 10^{3}$ & $(5.26 \pm 0.90) \times 10^{9}$ & $(1.31 \pm 0.40) \times 10^{0}$ & 75 \\ [1.5ex]
        WASP-11 & $4900\pm65$ & $7.6^{+6}_{-3.5}$ & $7\pm5$ & \nodata & $5.14\times10^{-2}$ & $(2.35 \pm 0.19) \times 10^{13}$ & $(4.70 \pm 0.38) \times 10^{3}$ & $(1.87 \pm 0.19) \times 10^{9}$ & $(2.46 \pm 1.56) \times 10^{-1}$ & 35,76,77,78 \\ [1.5ex]
        WASP-12 & $6313\pm52$ & $1.7\pm0.8$ & $59^{+15}_{-20}$ & \nodata & $3.63\times10^{-4}$ & $(4.60 \pm 0.45) \times 10^{12}$ & $(9.19 \pm 0.90) \times 10^{2}$ & $(2.47 \pm 0.34) \times 10^{7}$ & $(1.45 \pm 0.71) \times 10^{-2}$ & 17,79 \\ [1.5ex]
        WASP-13 & $6025\pm21$ & $7.4\pm0.4$ & $8^{+13}_{-12}$ & \nodata & $2.94\times10^{-2}$ & $(2.16 \pm 0.39) \times 10^{12}$ & $(4.33 \pm 0.79) \times 10^{2}$ & $(3.02 \pm 0.69) \times 10^{8}$ & $(4.08 \pm 0.96) \times 10^{-2}$ & 80,7,81 \\ [1.5ex]
        WASP-14 & $6462\pm75$ & $0.75\pm0.25$ & $-33.1\pm7.4$ & \nodata & $3.25\times10^{-5}$ & $(5.98 \pm 3.96) \times 10^{12}$ & $(1.20 \pm 0.79) \times 10^{3}$ & $(1.21 \pm 0.62) \times 10^{8}$ & $(1.61 \pm 0.99) \times 10^{-1}$ & 82,7 \\ [1.5ex]
        WASP-15 & $6573\pm70$ & $3.9^{+2.8}_{-1.3}$ & $-139.6^{+4.3}_{-5.2}$ & \nodata & $^{\mathrm{c}}5.83\times10^{-4}$ & $(3.51 \pm 0.24) \times 10^{16}$ & $(7.01 \pm 0.48) \times 10^{6}$ & $^{\mathrm{c}}(1.38 \pm 0.13) \times 10^{10}$ & $(3.53 \pm 1.89) \times 10^{0}$ & 68,83 \\ [1.5ex]
        WASP-16 & $5630\pm70$ & $2.3^{+5.8}_{-2.2}$ & $-4.2^{+11}_{-13.9}$ & \nodata & $3.05\times10^{-2}$ & $(1.14 \pm 0.10) \times 10^{12}$ & $(2.27 \pm 0.20) \times 10^{2}$ & $(1.52 \pm 0.21) \times 10^{8}$ & $(6.63 \pm 11.60) \times 10^{-2}$ & 84,83 \\ [1.5ex]
        WASP-17 & $6550\pm100$ & $3^{+0.9}_{-2.6}$ & $-148.5^{+4.2}_{-5.4}$ & \nodata & $1.14\times10^{-4}$ & $(3.57 \pm 0.51) \times 10^{16}$ & $(7.15 \pm 1.02) \times 10^{6}$ & $(8.04 \pm 1.06) \times 10^{10}$ & $(2.68 \pm 1.60) \times 10^{1}$ & 68,85 \\ [1.5ex]
        WASP-18 & $6400\pm70$ & $1\pm0.5$ & $4\pm5$ & \nodata & $2.20\times10^{-4}$ & $(2.22 \pm 1.06) \times 10^{11}$ & $(4.44 \pm 2.12) \times 10^{1}$ & $(1.44 \pm 0.52) \times 10^{6}$ & $(1.44 \pm 0.88) \times 10^{-3}$ & 68,86 \\ [1.5ex]
        WASP-19 & $5460\pm90$ & $11.5^{+2.7}_{-2.3}$ & $1\pm1.2$ & \nodata & $3.94\times10^{-2}$ & $(3.05 \pm 0.22) \times 10^{9}$ & $(6.09 \pm 0.43) \times 10^{-1}$ & $(3.17 \pm 0.28) \times 10^{5}$ & $(2.75 \pm 0.64) \times 10^{-5}$ & 87,88 \\ [1.5ex]
        WASP-20 & $6000\pm100$ & $7^{+2}_{-1}$ & $8.1\pm3.6$ & \nodata & $^{\mathrm{d}}1.94\times10^{-3}$ & $(2.51 \pm 0.21) \times 10^{13}$ & $(5.01 \pm 0.43) \times 10^{3}$ & $^{\mathrm{d}}(5.29 \pm 0.62) \times 10^{10}$ & $(7.56 \pm 1.85) \times 10^{0}$ & 89 \\ [1.5ex]
        WASP-22 & $6020\pm50$ & $3\pm1$ & $22\pm16$ & \nodata & $7.75\times10^{-3}$ & $(3.08 \pm 0.15) \times 10^{12}$ & $(6.16 \pm 0.30) \times 10^{2}$ & $(1.63 \pm 0.16) \times 10^{9}$ & $(5.43 \pm 1.89) \times 10^{-1}$ & 90 \\ [1.5ex]
        WASP-24 & $6297\pm58$ & $3.8^{+1.3}_{-1.2}$ & $-4.7\pm4$ & \nodata & $5.33\times10^{-3}$ & $(1.06 \pm 0.07) \times 10^{15}$ & $(2.12 \pm 0.13) \times 10^{5}$ & $(8.29 \pm 0.62) \times 10^{7}$ & $(2.18 \pm 0.74) \times 10^{-2}$ & 91,92 \\ [1.5ex]
        WASP-25 & $5736\pm35$ & $0.1^{+5.7}_{-0.1}$ & $14.6\pm6.7$ & \nodata & $2.91\times10^{-2}$ & $(1.45 \pm 0.17) \times 10^{13}$ & $(2.90 \pm 0.33) \times 10^{3}$ & $(2.04 \pm 0.32) \times 10^{9}$ & $(2.04 \pm 59.20) \times 10^{1}$ & 84,93 \\ [1.5ex]
        WASP-26 & $6034\pm31$ & $6\pm2$ & $-34^{+36}_{-26}$ & \nodata & $1.24\times10^{-2}$ & $(2.60 \pm 0.11) \times 10^{11}$ & $(5.20 \pm 0.21) \times 10^{1}$ & $(8.59 \pm 0.89) \times 10^{7}$ & $(1.43 \pm 0.50) \times 10^{-2}$ & 17,94 \\ [1.5ex]
        WASP-28 & $6100\pm150$ & $5^{+3}_{-2}$ & $8\pm18$ & \nodata & $3.74\times10^{-3}$ & $(1.56 \pm 0.14) \times 10^{12}$ & $(3.11 \pm 0.28) \times 10^{2}$ & $(1.70 \pm 0.21) \times 10^{9}$ & $(3.41 \pm 1.75) \times 10^{-1}$ & 89 \\ [1.5ex]
        WASP-30 & $6190\pm50$ & $1.5\pm0.5$ & $-7^{+27}_{-19}$ & \nodata & $8.61\times10^{-4}$ & $(4.22 \pm 5.74) \times 10^{8}$ & $(8.44 \pm 11.50) \times 10^{-2}$ & $(2.01 \pm 2.73) \times 10^{6}$ & $(1.34 \pm 1.87) \times 10^{-3}$ & 95 \\ [1.5ex]
        WASP-31 & $6175\pm70$ & $1^{+3}_{-0.5}$ & $2.8\pm3.1$ & \nodata & $3.02\times10^{-4}$ & $(4.15 \pm 0.39) \times 10^{12}$ & $(8.31 \pm 0.78) \times 10^{2}$ & $(5.63 \pm 0.64) \times 10^{10}$ & $(5.63 \pm 9.88) \times 10^{1}$ & 84,96 \\ [1.5ex]
        WASP-32 & $6100\pm100$ & $2.42^{+0.53}_{-0.56}$ & $10.5^{+6.4}_{-6.5}$ & \nodata & $1.08\times10^{-3}$ & $(4.93 \pm 0.47) \times 10^{10}$ & $(9.87 \pm 0.95) \times 10^{0}$ & $(1.87 \pm 0.27) \times 10^{8}$ & $(7.73 \pm 2.08) \times 10^{-2}$ & 30,97,84 \\ [1.5ex]
        WASP-33 & $7430\pm100$ & $0.1^{+0.4}_{-0.09}$ & $252\pm2$ & $103^{+5}_{-4}$ ($0.7^{+1.5}_{-1.6}\deg\,yr^{-1}$) & $1.00\times10^{-8}$ & $(5.01 \pm 4.95) \times 10^{12}$ & $(1.00 \pm 0.99) \times 10^{3}$ & $(6.32 \pm 4.62) \times 10^{11}$ & $(6.32 \pm 16.20) \times 10^{3}$ & 98,99,100,120 \\ [1.5ex]
        WASP-38 & $6150\pm80$ & $3.41^{+0.26}_{-0.24}$ & $7.5^{+4.7}_{-6.1}$ & \nodata & $2.86\times10^{-4}$ & $(1.72 \pm 0.13) \times 10^{12}$ & $(3.44 \pm 0.26) \times 10^{2}$ & $(2.46 \pm 0.23) \times 10^{10}$ & $(7.23 \pm 0.86) \times 10^{0}$ & 30,101,84 \\ [1.5ex]
        WASP-52 & $5000\pm100$ & $0.4^{+0.3}_{-0.2}$ & $24^{+17}_{-9}$ & \nodata & $5.53\times10^{-2}$ & $(1.57 \pm 0.11) \times 10^{12}$ & $(3.15 \pm 0.22) \times 10^{2}$ & $(1.17 \pm 0.11) \times 10^{8}$ & $(2.92 \pm 1.84) \times 10^{-1}$ & 102 \\ [1.5ex]
        WASP-66 & $6600\pm150$ & $3.3^{+10}_{-2.7}$ & $-4\pm22$ & \nodata & $1.30\times10^{-7}$ & $(1.11 \pm 0.25) \times 10^{15}$ & $(2.22 \pm 0.49) \times 10^{5}$ & $(2.41 \pm 0.51) \times 10^{12}$ & $(7.31 \pm 14.10) \times 10^{2}$ & 103,104 \\ [1.5ex]
        WASP-71 & $6180\pm52$ & $2.5\pm0.5$ & $19.8\pm9.9$ & \nodata & $^{\mathrm{e}}5.27\times10^{-4}$ & $(9.31 \pm 0.87) \times 10^{9}$ & $(1.86 \pm 0.17) \times 10^{0}$ & $^{\mathrm{e}}(7.24 \pm 1.50) \times 10^{7}$ & $(2.90 \pm 0.84) \times 10^{-2}$ & 105 \\ [1.5ex]
        WASP-79 & $6600\pm100$ & $0.5$ & $-106^{+19}_{-13}$ & \nodata & $3.40\times10^{-5}$ & $(1.65 \pm 0.37) \times 10^{16}$ & $(3.30 \pm 0.74) \times 10^{6}$ & $(1.21 \pm 0.25) \times 10^{11}$ & $(2.41 \pm 0.50) \times 10^{2}$ & 106,107 \\ [1.5ex]
        WASP-80 & $4145\pm100$ & $0.1^{+0.03}_{-0.02}$ & $75\pm4$ & \nodata & $7.09\times10^{-2}$ & $(1.21 \pm 0.11) \times 10^{13}$ & $(2.42 \pm 0.23) \times 10^{3}$ & $(7.00 \pm 0.81) \times 10^{8}$ & $(7.00 \pm 1.93) \times 10^{0}$ & 108,109 \\ [1.5ex]
        WASP-87 & $6450\pm110$ & $3.8\pm0.8$ & $-8\pm11$ & \nodata & $6.00\times10^{-8}$ & $(1.05 \pm 0.29) \times 10^{13}$ & $(2.11 \pm 0.58) \times 10^{3}$ & $(1.94 \pm 0.45) \times 10^{11}$ & $(5.10 \pm 1.59) \times 10^{1}$ & 103,110 \\ [1.5ex]
        \hline \\ [-2.0ex]
    \end{tabular}}
\end{table*}

\hfill
\begin{table*}
  \centering
  \resizebox{\textwidth}{!}{%
    \begin{threeparttable}[b]
      \begin{tabular}{l c c c c c c c c c c}
        \multicolumn{11}{c}{\large \textbf{Table 6} -- \emph{Continued}}\\ [2ex]
        \hline\hline \\ [-2.0ex]
        \multirow{2}{*}{System} &
        $T_{eff}$ &
        Age &
        $\lambda$ &
        $\psi$ &
        $M_{cz}$ &
        $\tau_{\mathrm{RA\, or\, CE}}$ &
        \multirow{2}{*}{$\frac{\tau_{\mathrm{RA\, or\, CE}}(\mathrm{year})}{5\times10^{9}(\mathrm{year})}$} & 
        $\tau_{mcz}$ &
        \multirow{2}{*}{$\frac{\tau_{mcz}(\mathrm{year})}{\mathrm{Age}(\mathrm{year})}$} &
        \multirow{2}{*}{References} \\
        & (K) & (Gyr) & (deg) & (deg) & ($M_{\odot}$) & (year) &  & (year) &  & \\ [0.5ex]
        \hline \\ [-2.0ex]
        WASP-94Ab & $6170\pm80$ & $4$ & $151^{+16}_{-23}$ & \nodata & $8.15\times10^{-5}$ & $(4.16 \pm 0.46) \times 10^{12}$ & $(8.33 \pm 0.92) \times 10^{2}$ & $(2.09 \pm 0.29) \times 10^{11}$ & $(5.23 \pm 0.72) \times 10^{1}$ & 111 \\ [1.5ex]
        WASP-103 & $6110\pm160$ & $4$ & $3\pm33$ & \nodata & $1.63\times10^{-3}$ & $(1.24 \pm 0.16) \times 10^{9}$ & $(2.47 \pm 0.33) \times 10^{-1}$ & $(3.11 \pm 0.47) \times 10^{6}$ & $(7.77 \pm 1.18) \times 10^{-4}$ & 103,112 \\ [1.5ex]
        WASP-111 & $6400\pm150$ & $2.6\pm0.6$ & $-5\pm16$ & \nodata & $1.30\times10^{-7}$ & $(8.73 \pm 2.18) \times 10^{13}$ & $(1.75 \pm 0.44) \times 10^{4}$ & $(5.02 \pm 1.17) \times 10^{11}$ & $(1.93 \pm 0.63) \times 10^{2}$ & 110 \\ [1.5ex]
        WASP-117 & $6040\pm90$ & $4.6\pm2$ & $-44\pm11$ & $69.6^{+4.7}_{-4.1}$ & $3.67\times10^{-4}$ & $(1.23 \pm 0.06) \times 10^{15}$ & $(2.46 \pm 0.13) \times 10^{5}$ & $(1.38 \pm 0.20) \times 10^{13}$ & $(2.99 \pm 1.37) \times 10^{3}$ & 113 \\ [1.5ex]
        WASP-121 & $6460\pm140$ & $1.5\pm1$ & $257.8^{+5.3}_{-5.5}$ & \nodata & $2.15\times10^{-5}$ & $(3.30 \pm 0.45) \times 10^{13}$ & $(6.61 \pm 0.90) \times 10^{3}$ & $(1.86 \pm 0.23) \times 10^{9}$ & $(1.24 \pm 0.84) \times 10^{0}$ & 114 \\ [1.5ex]
        XO-2 & $5340\pm50$ & $5.3^{+1}_{-0.7}$ & $10\pm72$ & \nodata & $3.57\times10^{-2}$ & $(1.85 \pm 0.14) \times 10^{12}$ & $(3.69 \pm 0.28) \times 10^{2}$ & $(2.12 \pm 0.31) \times 10^{8}$ & $(4.00 \pm 0.86) \times 10^{-2}$ & 115,116 \\ [1.5ex]
        XO-3 & $6429\pm75$ & $2.82^{+0.58}_{-0.82}$ & $37.3\pm3$ & \nodata & $8.61\times10^{-5}$ & $(4.69 \pm 2.85) \times 10^{13}$ & $(9.39 \pm 5.70) \times 10^{3}$ & $(1.44 \pm 0.66) \times 10^{8}$ & $(5.11 \pm 2.67) \times 10^{-2}$ & 117,15 \\ [1.5ex]
        XO-4 & $6397\pm70$ & $2.1\pm0.6$ & $-46.7^{+8.1}_{-6.1}$ & \nodata & $3.76\times10^{-6}$ & $(7.75 \pm 1.45) \times 10^{15}$ & $(1.55 \pm 0.29) \times 10^{6}$ & $(4.15 \pm 1.61) \times 10^{11}$ & $(1.98 \pm 0.95) \times 10^{2}$ & 118,5 \\ [1.5ex]
        \hline \\ [-2.0ex]
      \end{tabular}
      \normalsize
      \textbf{Note} -- $T_\mathrm{eff}$ is the effective temperature of the host star, $\lambda$ is the projected orbital obliquity, $\psi$ is the true orbital obliquity, $M_{cz}$ is the mass in the convective zone as described in   \autoref{sec:Discussion_wasp-66-103}, $\tau_{RA}$ and $\tau_{CE}$ are the radiative and convective timescale for alignment, respectively, and $\tau_{mcz}$ is the timescale when considering the mass in the convective zone.  $\tau_{RA}$ and $\tau_{CE}$ timescales have also been normalized by dividing by $5\times10^{9}$. $\tau_{mcz}$ timescale is also normalized to the age of the system. The sample of planetary systems with measured spin--orbit angles were selected from the \href{http://www.astro.keele.ac.uk/jkt/tepcat/tepcat.html}{\it Rossiter-McLaughlin Catalogue} and the \href{http://www.astro.physik.uni-goettingen.de/~rheller/}{\it Ren\'{e} Heller's Holt-Rossiter-McLaughlin Encyclopaedia} on 2015 November. $\lambda$ and $\psi$ values, along with the associated uncertainties were obtained from the \textit{Holt-Rossiter-McLaughlin Encyclopaedia}. The other stellar and planetary parameters used to calculate $M_{cz}$, as well as the realignment timescales ($\tau_{\mathrm{RA}}$, $\tau_{\mathrm{CE}}$, and $\tau_{mcz}$) were taken from \textit{The Rossiter-McLaughlin Catalogue}, \textit{The Extrasolar Planets Encyclopaedia}, and the references provided. \\ [0.75ex]
      \textbf{$^{\mathrm{a}}$} Systems with no reported age in the literature are set to 0.5\,Gyr. \\ [0.75ex]
      \textbf{$^{\mathrm{b}}$} We found HAT-P-13 had started evolving off the main-sequence at $\sim3.5$\,Gyr and was quite evolved at the reported age of $5.0$\,Gyr, using the EZ-Web stellar evolution tool. Therefore, we determined $M_{cz}$ and computed $\tau_{mcz}$ assuming a stellar age at $3.5$\,Gyr. \\ [0.75ex]
      \textbf{$^{\mathrm{c}}$} We found WASP-15 to be pretty evolved at the reported age of $3.9$\,Gyr, using the EZ-Web stellar evolution tool. Therefore, we set the stellar age to the lower age estimate of $2.6$\,Gyr to determine $M_{cz}$ and compute $\tau_{mcz}$. \\ [0.75ex]
      \textbf{$^{\mathrm{d}}$} The EZ-Web tool stopped evolving WASP-20 after $\sim6$\,Gyr since it had evolved well off the main-sequence and is likely a white dwarf at the reported age of $7$\,Gyr. We thus assumed a stellar age of $1$\,Gyr for determining $M_{cz}$ and computing $\tau_{mcz}$ since the stellar properties from EZ-Web best matched those that are published. \\ [0.75ex]
      \textbf{$^{\mathrm{e}}$} We found WASP-71 to be fairly evolved at the reported age of $2.5$\,Gyr, using the EZ-Web stellar evolution tool. Therefore, we set the stellar age to the lower age estimate of $2.0$\,Gyr to determine $M_{cz}$ and compute $\tau_{mcz}$.  \\ [1.5ex]
      \textbf{References} -- The references are taken from the \href{http://www.astro.physik.uni-goettingen.de/~rheller/}{\it Ren\'{e} Heller's Holt-Rossiter-McLaughlin Encyclopaedia}, \href{http://www.astro.keele.ac.uk/jkt/tepcat/tepcat.html}{\it Rossiter-McLaughlin Catalogue}, and the \href{http://exoplanet.eu/}{\it Extrasolar Planets Encyclopaedia} and are as follows;
      $(1)$: \citealt{2014A&A...569A..65B};
      $(2)$: \citealt{2011ApJ...740...49V};
      $(3)$: \citealt{2012A&A...539A..28G};
      $(4)$: \citealt{2010MNRAS.402L...1P};
      $(5)$: \citealt{2011MNRAS.417.2166S};
      $(6)$: \citealt{2008A&A...482L..25B};
      $(7)$: \citealt{2012MNRAS.426.1291S};
      $(8)$: \citealt{2009A&A...506..377T};
      $(9)$: \citealt{2012A&A...543L...5G};
      $(10)$: \citealt{2011A&A...533A.130H};
      $(11)$: \citealt{2012A&A...537A.136G};
      $(12)$: \citealt{2008ApJ...686..649J};
      $(13)$: \citealt{2014MNRAS.437...46N};
      $(14)$: \citealt{2008A&A...481..529L};
      $(15)$: \citealt{2010MNRAS.408.1689S};
      $(16)$: \citealt{2011AJ....141...63W};
      $(17)$: \citealt{2012ApJ...757...18A};
      $(18)$: \citealt{2012MNRAS.422.3099S};
      $(19)$: \citealt{2011A&A...533A.113M};
      $(20)$: \citealt{2013A&A...551A..11M};
      $(21)$: \citealt{2010ApJ...723L.223W};
      $(22)$: \citealt{2010ApJ...718..575W};
      $(23)$: \citealt{2012MNRAS.420.2580S};
      $(24)$: \citealt{2013A&A...557A..30C};
      $(25)$: \citealt{2013ApJ...772...80F};
      $(26)$: \citealt{2012ApJ...749..134H};
      $(27)$: \citealt{2014A&A...564L..13E};
      $(28)$: \citealt{2011ApJ...742..116B};
      $(29)$: \citealt{2010ApJ...725.2017K};
      $(30)$: \citealt{2012ApJ...760..139B};
      $(31)$: \citealt{2011ApJ...734..109B};
      $(32)$: \citealt{2011ApJ...735...24J};
      $(33)$: \citealt{2011ApJ...742...59H};
      $(34)$: \citealt{2012AJ....144...19B};
      $(35)$: \citealt{2015A&A...579A.136M};
      $(36)$: \citealt{2013A&A...558A..55M};
      $(37)$: \citealt{2014ApJ...792..112A};
      $(38)$: \citealt{2013AJ....146..113B};
      $(39)$: \citealt{2015ApJ...814L..16Z};
      $(40)$: \citealt{2009PASJ...61..991N};
      $(41)$: \citealt{2010A&A...516A..95H};
      $(42)$: \citealt{2009ApJ...703.2091W};
      $(43)$: \citealt{2009ApJ...696..241C};
      $(44)$: \citealt{2010MNRAS.403..151C};
      $(45)$: \citealt{2005ApJ...631.1215W};
      $(46)$: \citealt{2012ApJ...761..123S};
      $(47)$: \citealt{2014ApJ...790...30J};
      $(48)$: \citealt{2014ApJ...788...92S};
      $(49)$: \citealt{2011ApJ...736L...4S};
      $(50)$: \citealt{2011ApJS..197...14D};
      $(51)$: \citealt{2013ApJ...771...11A};
      $(52)$: \citealt{2014ApJS..210...20M};
      $(53)$: \citealt{2012Natur.487..449S};
      $(54)$: \citealt{2013ApJ...766..101C};
      $(55)$: \citealt{2013Sci...342..331H};
      $(56)$: \citealt{2013ApJ...775...54S};
      $(57)$: \citealt{2012ApJ...759L..36H};
      $(58)$: \citealt{2013ApJ...768...14W};
      $(59)$: \citealt{2015A&A...579A..55B};
      $(60)$: \citealt{2014A&A...571A..37S};
      $(61)$: \citealt{2013A&A...554A..28C};
      $(62)$: \citealt{2007PASJ...59..763N};
      $(63)$: \citealt{2008ApJ...682.1283W};
      $(64)$: \citealt{2010PASJ...62..653N};
      $(65)$: \citealt{2011ApJ...738...50A};
      $(66)$: \citealt{2014AcA....64...27M};
      $(67)$: \citealt{2007MNRAS.379..773S};
      $(68)$: \citealt{2010A&A...524A..25T};
      $(69)$: \citealt{2010A&A...523A..52M};
      $(70)$: \citealt{2013AJ....146..147M};
      $(71)$: \citealt{2008MNRAS.385.1576P};
      $(72)$: \citealt{2009A&A...496..259G};
      $(73)$: \citealt{2009A&A...501..785G};
      $(74)$: \citealt{2012ApJ...744..189A};
      $(75)$: \citealt{2010A&A...517L...1Q};
      $(76)$: \citealt{2014AJ....147...92W};
      $(77)$: \citealt{2009ApJ...696.1950B};
      $(78)$: \citealt{2009A&A...502..395W};
      $(79)$: \citealt{2013MNRAS.436.2956S};
      $(80)$: \citealt{2014MNRAS.440.3392B};
      $(81)$: \citealt{2012MNRAS.419.1248B};
      $(82)$: \citealt{2009PASP..121.1104J};
      $(83)$: \citealt{2013MNRAS.434.1300S};
      $(84)$: \citealt{2012MNRAS.423.1503B};
      $(85)$: \citealt{2012MNRAS.426.1338S};
      $(86)$: \citealt{2013MNRAS.428.2645M};
      $(87)$: \citealt{2013MNRAS.428.3671T};
      $(88)$: \citealt{2013MNRAS.436....2M};
      $(89)$: \citealt{2015A&A...575A..61A};
      $(90)$: \citealt{2011A&A...534A..16A};
      $(91)$: \citealt{2011MNRAS.414.3023S};
      $(92)$: \citealt{2012A&A...545A..93S};
      $(93)$: \citealt{2011MNRAS.410.1631E};
      $(94)$: \citealt{2013MNRAS.432..693M};
      $(95)$: \citealt{2013A&A...549A..18T};
      $(96)$: \citealt{2011A&A...531A..60A};
      $(97)$: \citealt{2010PASP..122.1465M};
      $(98)$: \citealt{2010MNRAS.407..507C};
      $(99)$: \citealt{2013A&A...553A..44K};
      $(100)$: \citealt{2011A&A...535A.110M};
      $(101)$: \citealt{2011A&A...525A..54B};
      $(102)$: \citealt{2013A&A...549A.134H};
      $(103)$: This work;
      $(104)$: \citealt{2012MNRAS.426..739H};
      $(105)$: \citealt{2013A&A...552A.120S};
      $(106)$: \citealt{2013ApJ...774L...9A};
      $(107)$: \citealt{2012A&A...547A..61S};
      $(108)$: \citealt{2013A&A...551A..80T};
      $(109)$: \citealt{2014A&A...562A.126M};
      $(110)$: \citealt{2014arXiv1410.3449A};
      $(111)$: \citealt{2014A&A...572A..49N};
      $(112)$: \citealt{2014A&A...562L...3G};
      $(113)$: \citealt{2014A&A...568A..81L};
      $(114)$: \citealt{2015arXiv150602471D};
      $(115)$: \citealt{2011PASJ...63L..67N};
      $(116)$: \citealt{2012ApJ...761....7C};
      $(117)$: \citealt{2011PASJ...63L..57H};
      $(118)$: \citealt{2010PASJ...62L..61N};
      $(119)$: \citealt{2015ApJ...805...28M};
      $(120)$: \citealt{2016MNRAS.455..207I}.
    \end{threeparttable}}
\end{table*}
\hfill
\clearpage

\end{document}